%% file: ms_arXiv.tex
\documentclass[twocolumn]{aastex631}

\usepackage{listings}
\usepackage{amsmath}

\shorttitle{47 Tuc in Rubin DP1}
\shortauthors{Choi et al.}

\begin{document}

\title{47 Tuc in Rubin Data Preview 1: Exploring Early LSST Data and Science Potential}

\correspondingauthor{Yumi Choi}
\email{yumi.choi@noirlab.edu}

\input{authors}

\begin{abstract}
We present analyses of the early data from Rubin Observatory's Data Preview 1 (DP1) for the globular cluster 47 Tuc field. The DP1 dataset for 47 Tuc includes four nights of observations from the Rubin Commissioning Camera (LSSTComCam), covering multiple bands ($ugriy$). We address challenges of crowding in the inner region of the cluster and toward the SMC in DP1, and demonstrate improved star-galaxy separation by fitting fifth-degree polynomials to the stellar loci in color-color diagrams and applying multi-dimensional sigma clipping. We compile a catalog of 3,576 probable 47 Tuc member stars selected via a combination of isochrone, Gaia proper-motion, and color-color space matched filtering. We explore the sources of photometric scatter in the 47 Tuc color-color sequence, evaluating contributions from various potential sources, including differential extinction within the cluster. Finally, of the 72 well-characterized variables in the field, we recover five known variable stars, including three RR Lyrae and two eclipsing binaries, in the coadd-based object catalog, and identify 62 in the difference image-based object catalog. Although the DP1 lightcurves have sparse temporal sampling, they appear to follow the patterns of densely-sampled literature lightcurves well. Despite some data limitations for crowded-field stellar analysis, DP1 demonstrates the promising scientific potential for future LSST data releases.   
\end{abstract}


\section{Introduction} \label{sec:intro}
The Vera C. Rubin Observatory, a joint NSF and DOE initiative situated on Cerro Pach\'on in northern Chile, is designed to carry out the Legacy Survey of Space and Time \citep[LSST;][]{Ivezic2019}. The LSST is a decade-long program capturing rapid, deep, multi-band imaging of the entire southern sky at high spatial and temporal resolution across the six $ugrizy$ bands. Before full survey operations begin, commissioning efforts produce small, compared to the future official data releases, volumes of good quality data intended to test and refine the observatory's systems. To support early scientific investigations and community readiness, Rubin offers Data Previews through its Early Science Program \citep{RTN-011}, delivering preliminary but science-grade products and early access to data services. 

As the earliest on-sky campaign, the Rubin Commissioning Camera (LSSTComCam) was used to align and verify the end-to-end functionality of the Rubin Observatory's systems, resulting in 1,792 science-grade exposures collected in late 2024. These data were processed with the LSST Science Pipelines \citep{Bosch2019,PSTN-019} to produce Data Preview 1 \citep[DP1\footnote{\url{https://dp1.lsst.io}};][]{RTN-095}. The available DP1 data products comprise raw and calibrated single-epoch exposures, coadds, difference images, and various detection catalogs, totaling $\sim$3.5~TB. It contains $\sim$2.3 million distinct astronomical objects, and is accessible via the Rubin Science Platform, a cloud-based data science environment for data analysis. One of the DP1 target fields is centered on the globular cluster 47 Tucanae (47 Tuc). 

47~Tuc, aka NGC~104, is the second-brightest globular cluster in the sky, with $V \approx 4$ mag. Its distance has been measured with high precision to be $\sim$4.45~kpc using, for example, Hubble Space Telescope (HST) proper motion by \citet{Watkins2015} and Gaia DR2 parallax by \citet{Chen2018}, which corresponds to a true distance modulus of $13.24 \pm 0.06$.  The line-of-sight reddening toward this relatively high-latitude cluster was estimated at $E(B-V) \approx 0.03$ by \citet{Hesser1976} based on Str\"omgren photometry of distant early-type stars along that sightline. This low value, consistent with its Galactic latitude of $b = -44^\circ.8$,  has been propitious, making it possible for 47 Tuc to be used as a template for stellar populations in a metal rich globular cluster, whereas other metal rich ones are to be found in the Galactic plane and plagued with high reddening that varies on the scale of a cluster. 

Its HST and Gaia distance is in agreement with the distance modulus of $13.21 \pm 0.08$, derived from model analysis of a detached binary (V69) star in the cluster by \citet{Brogaard2017}.  Other prior values, such as from the modeling of color-magnitude diagram features, and the cooling curves of its white dwarfs are dependent on knowing or assuming the detailed metal content and the age of 47 Tuc's stars. Thus Gaia's parallax based distance sets the stage for a more sure-footed determination of the properties of the cluster's stellar content, its age, and its history. Despite being an early known metal rich stellar system (e.g., $[Fe/H] \approx -0.75$ \citep{Renno2020} or $[M/H] \approx -0.5$ \citep{Simunovic2023}), the morphology of its color-magnitude diagram implies an age of between $11$ and $13$~Gyrs (more specifically \citet{Thompson_2020} claim $12.0 \pm 0.5$~Gyr), making it as old (to within discernible uncertainties) as the oldest globular clusters of the Galaxy. It has been studied using the so called ``chromosome maps", which trace abundance correlations among stellar populations and reveals that its oldest stars likely followed multiple evolutionary and chemical enrichment pathways \citep[e.g.,][]{Carretta2005, Milone2012, Marino_2023}.   Given its proximity, it also has been a target for studying the cooling sequence of its white dwarfs. Since 47 Tuc is a well-studied cluster, the parameter values listed here are illustrative examples rather than exhaustive or unique measurements. 

On the sky, 47 Tuc sits in front of the western extension of the Small Magellanic Cloud (SMC).  A concentration of stars situated $\sim$15$\arcmin$ from the center of 47 Tuc was first reported by \citet{Bellazini2005} and surmised to be a background star cluster (Bologna A) and later confirmed to be an SMC cluster \citep{Bellazzini2019}.  47 Tuc itself has a highly concentrated core (Shapley class III), with claims that it contains an intermediate mass black hole \citep{DellaCroce_2024}.  By contrast, it also  exhibits an extended periphery of stars: e.g.  \citet{Piatti_2017} has claimed to find stars out to 5.5 times the tidal radius of $\sim42\arcmin$, that indicate a halo like structure as seen in NGC~1851 \citep{Olszewski_2009}. There is a radial gradient in the stellar population \citep[e.g.,][]{Li_2014}, involving metallicities, helium abundance, and mass segregation.  These characteristics collectively feed the speculation that like $\omega$~Cen and M54, 47~Tuc is the surviving central remnant from a dwarf galaxy merger.

The addition of proper motions from Gaia for 47 Tuc completes the 6-dimensional phase space characterization of this cluster, providing the observational prerequisites for good dynamical models of its motion through the Galaxy \citep[e.g.,][]{Katz2023}. LSST will build on this foundation by delivering deeper, wider, and more time-resolved imaging, extending precise photometry and proper motions to far fainter magnitudes than Gaia can reach \citep{Ivezic2019}, improving constraints on the cluster's internal and orbital kinematics. Such data are also essential for tracing 47 Tuc's stellar populations to the largest possible radii and for cleanly separating its members from multiple sources of contamination: background stars from the SMC, ambient ``stellar field'' of the Galaxy through which 47~Tuc is passing, and any debris potentially stripped from a possible dwarf galaxy progenitor of 47 Tuc. DP1 provides an early look at real data from the Rubin Observatory systems which can inform us of the strengths and limitations of the data, as well as the state of our currently available methodology for attaining the required precision and accuracy in photometry and proper motions for disambiguating the stellar population components.

In this paper, we use the 47 Tuc field to evaluate DP1 performance in a particularly challenging regime and to demonstrate its early science potential. We aim to (1) showcase key DP1 data products for this field, (2) develop and apply methods for robust member-star selection and star-galaxy separation, (3) assess photometric precision, saturation limits, and sources of scatter, and (4) examine the recovery of known variable stars.
   
This paper is organized as follows. In Section~\ref{sec:data}, we briefly describe the observation and the DP1 data products for the 47 Tuc field, and then present the field-specific data properties, such as crowding,  deblending, and star-galaxy separation. Section~\ref{sec:analysis} describes the 47 Tuc member star selection process, explores the photometry of selected member stars, and investigates the LSSTComCam detection of known variable stars. We summarize our main findings and briefly discuss potential with future data releases in Section~\ref{sec:summary}.

%====== Figure
\begin{figure*}
 \centering
      \includegraphics[width=\linewidth]{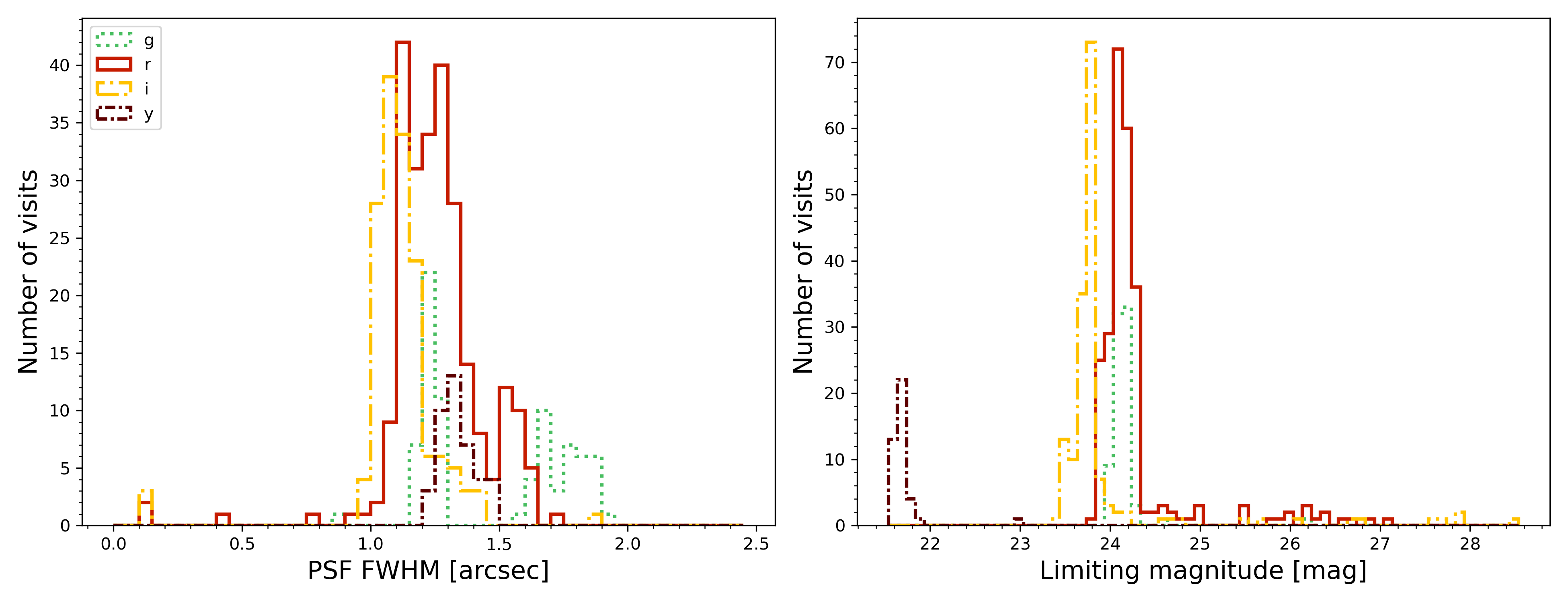}
       \caption{(Left:) Histogram of delivered image quality (PSF FWHM) for individual detector images in the $g$, $r$, $i$, and $y$ bands. Most observations fall between 0.8 and 2.0 arcseconds, with all the bands being the most frequently observed between 1.1 and 1.3\arcsec. (Right:) Histogram of 5$\sigma$ limiting magnitudes per detector image, also separated by band, with the deepest exposures in the $r$ and $g$ bands.
      \label{fig:psf_limitmags}}
\end{figure*}

\begin{deluxetable*}{lcccccc}
\tablecaption{Observing statistics, conditions, and areal coverage per Filter. \label{tab:47tuc_filters}}
\tablehead{
\colhead{Property} & \colhead{$u$} & \colhead{$g$} & \colhead{$r$} & \colhead{$i$} & \colhead{$z$} & \colhead{$y$}
}
\startdata
$N_{\mathrm{visits}}$\tablenotemark{a} & 6 (0) & 10 (10) & 33 (32) & 19 (19) & \dots & 5 (5) \\
Observing Nights & \dots &
\shortstack{2024-12-01\\2024-12-07} &
\shortstack{2024-11-25\\2024-11-26\\2024-12-01\\2024-12-07} &
\shortstack{2024-11-25\\2024-11-26\\2024-12-07} &
\dots & 2024-11-26 \\
$N_{\mathrm{CCDs}}$\tablenotemark{b} & \dots & 90 & 288 & 171 & \dots & 45 \\
Median Seeing ($''$) & \dots & 1.27 & 1.25 & 1.11 & \dots & 1.33 \\
Median $5\sigma$ Limiting Mag & \dots & 24.14 & 24.14 & 23.77 & \dots & 21.66 \\
Maximum $5\sigma$ Limiting Mag & \dots & 26.15 & 27.07 & 28.63 & \dots & 23.03 \\
Areal Coverage (deg$^{2}$)\tablenotemark{c} & \dots & 0.980 (0.825) & 1.064 (0.915) & 1.166 (0.997) & \dots & 0.807 (0.747) \\
\enddata
\tablenotetext{a}{Numbers in parentheses indicate the subset of visits included in the DP1 final processing.}
\tablenotetext{b}{Number of CCDs contributing to CCD-level statistics per filter.}
\tablenotetext{c}{Numbers in parentheses indicate the areal coverage with the central hole excluded.}
\end{deluxetable*}

\section{Data for the 47 Tuc Field} 
In this section, we summarize the relevant DP1 observations for the 47 Tuc field, briefly describe the processing and data products, and highlight field-specific characteristics, such as depth, coverage, crowding, and star-galaxy separation performance.

\subsection{Observations}\label{sec:observation}
The 47 Tuc field was one of seven target fields observed during the LSSTComCam \citep[the Rubin Observatory Commissioning Camera;][]{comcam} on-sky campaign (24 October--11 December 2024), and is centered at (RA, Dec) = (6.128, –72.090) degrees. Observations were obtained in five bands ($ugriy$), with the deepest coverage in the $r$ band. Imaging employed random dithers within a 0\fdg2 radius and occasional rotator offsets. The dense stellar environment of 47 Tuc makes it a valuable testbed for wavefront sensing, internal calibrations, and crowded-field stellar population studies. Although saturation in the cluster core led to missing coadds and prevented reliable measurements in that region, this target field still offers a great dataset for scientific analysis. The LSSTComCam observations for the 47 Tuc field include 6 visits in the $u$ band, 10 in $g$, 33 in $r$, 19 in $i$, and 5 in $y$. However, all $u$-band visits and one $r$-band visit failed to meet the quality criteria and were excluded from the final DP1 processing. Details about the LSSTComCam on-sky campaign and the DP1 can be found in SITCOM-149 \citep{SITCOMTN-149} and DP1 paper \citep{RTN-095}. 

In Figure~\ref{fig:psf_limitmags}, we present the distributions of the delivered image quality and depth across individual CCD images for the $g$, $r$, $i$, and $y$ bands. The left panel shows the distribution of PSF FWHM, with most images having seeing between 0.8\arcsec and 2.0\arcsec, and a peak around 1.1-1.3\arcsec across all bands. The right panel shows the distribution of 5~$\sigma$ limiting magnitudes, with the median depths reaching approximately 24.14, 24.14, 23.77, and 21.66~mag in the $g$, $r$, $i$, and $y$ bands, respectively. A per-CCD summary of observational properties by filter is listed in Table~\ref{tab:47tuc_filters}. 

We note that 47 Tuc lies within the footprint of the Wide-Fast-Deep (WFD) component of the LSST\footnote{https://survey-strategy.lsst.io/baseline/wfd.html}. According to the current baseline strategy, each WFD field is expected to receive approximately 800 visits over the 10-year survey with the full LSSTCam \citep{Lange2024,Roodman2024}. While this DP1 dataset offers a valuable early look at Rubin performance in crowded fields, the final LSST dataset will deliver an order of magnitude more visits and significantly deeper coadds, improving both the photometric depth and temporal sampling on 47 Tuc.

\subsection{DP1 Data Products for 47 Tuc}\label{sec:data}
All data were processed with the LSST Science Pipelines software \citep{Bosch2019, PSTN-019} at the U.S. Data Facility (USDF; housed at the SLAC National Laboratory). Initial processing steps in instrument signature removal (ISR) include applying bias, dark, and flat-field frames, corrections for crosstalk and the brighter-fatter effect \citep{Broughton2024}, and removal of other instrumental artifacts and effects (see \citealt{Plazas2025} for details about the ISR procedures, and \citealt{SITCOMTN-086} for more information on the calibration data products). Source detection and deblending is then run on the ISR-corrected images, followed by measurements of source centroids, shapes, and photometry. Initial photometry and shape estimates are used for the selection of stars for point-spread function (PSF) fitting; a preliminary fit to the PSF is then determined using a modified version of \texttt{PSFEx} \citep{Bertin2011}, after which a second, more careful and accurate, round of PSF fitting is performed using a modified version of \texttt{Piff} \citep{Jarvis2021a, Jarvis2021b}. Initial photometric and astrometric calibrations were performed by cross-matching to the Monster reference catalog -- an amalgam based on Gaia DR3 astrometry and photometry, and also including transformed DES, PanSTARRS1, SkyMapper, VST, and SDSS photometry \citep{DMTN-277}. Subsequently, more detailed determination of the astrometric solution uses fitting routines from \texttt{gbdes} \citep{Bernstein2017, Bernstein2022}, and a global photometric calibration is performed using the Forward Global Calibration Method (FGCM; \citealt{Burke2018}). Additional steps such as background estimation and subtraction are performed, and detection and measurement steps are repeated on the final per-detector \texttt{visit\_image}s produced by the single-visit processing.

In full data release processing, all \texttt{visit\_image} datasets are resampled onto a common sky projection (defined by a skymap\footnote{A skymap is a tiling of the celestial sphere that divides large-scale sky coverage into manageable sections for processing and analysis.}), convolved to a common PSF, and then stacked (with inverse-variance weighting) into a ``coadd'' image. Outlier rejection is performed to prevent image artifacts, cosmic rays, and moving objects from printing through to the final coadd images. \texttt{Object} catalogs \citep{Objecttable} are produced from detection and measurement of objects on the final \texttt{deep\_coadd} images \citep{deepcoadd}.

Details of the data products released with Data Preview 1 \citep[DP1;][]{dp1datasets} are provided in \url{https://dp1.lsst.io}. Briefly, the DP1 data products include \texttt{raw} images, processed per-detector single-frame \texttt{visit\_image}s and their corresponding \texttt{Source} catalogs, and stacked \texttt{deep\_coadd} images and their corresponding \texttt{Object} catalogs, among others.

%====== Figure
\begin{figure}
 \centering
      \includegraphics[width=\linewidth]{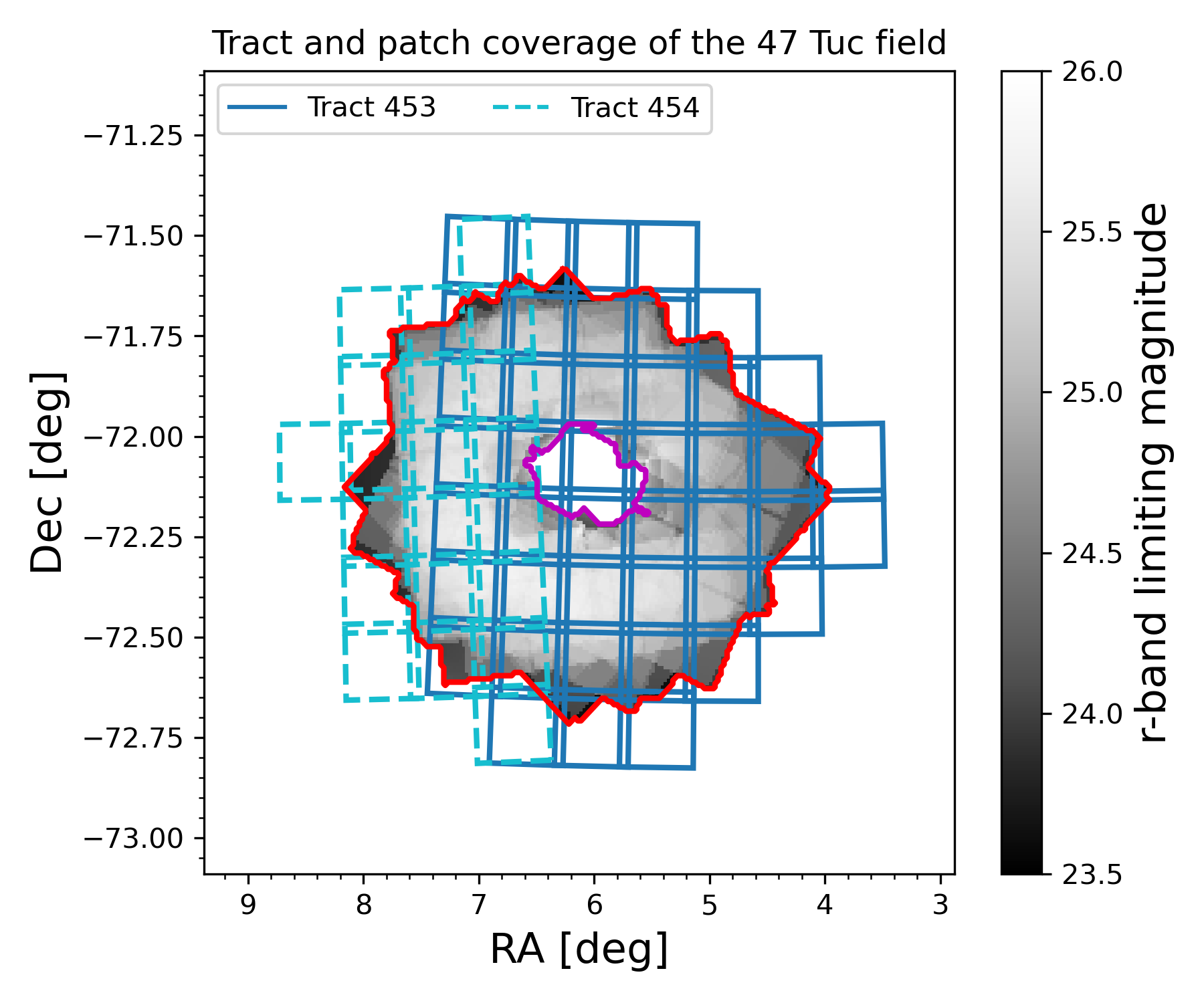}
       \caption{$r$-band 5$\sigma$ limiting magnitude map for the 47 Tuc field, shown in greyscale. Patch boundaries are overplotted and color-coded by tract, as indicated in the legend. The core of 47 Tuc is saturated in the LSSTComCam commissioning images, resulting in missing coadds in that region. This map is created purely based on cumulative images assuming perfect detection and measurement. 
      \label{fig:depthcoverage}}
\end{figure}

%====== Figure
\begin{figure*}
 \centering
      \includegraphics[width=\linewidth]{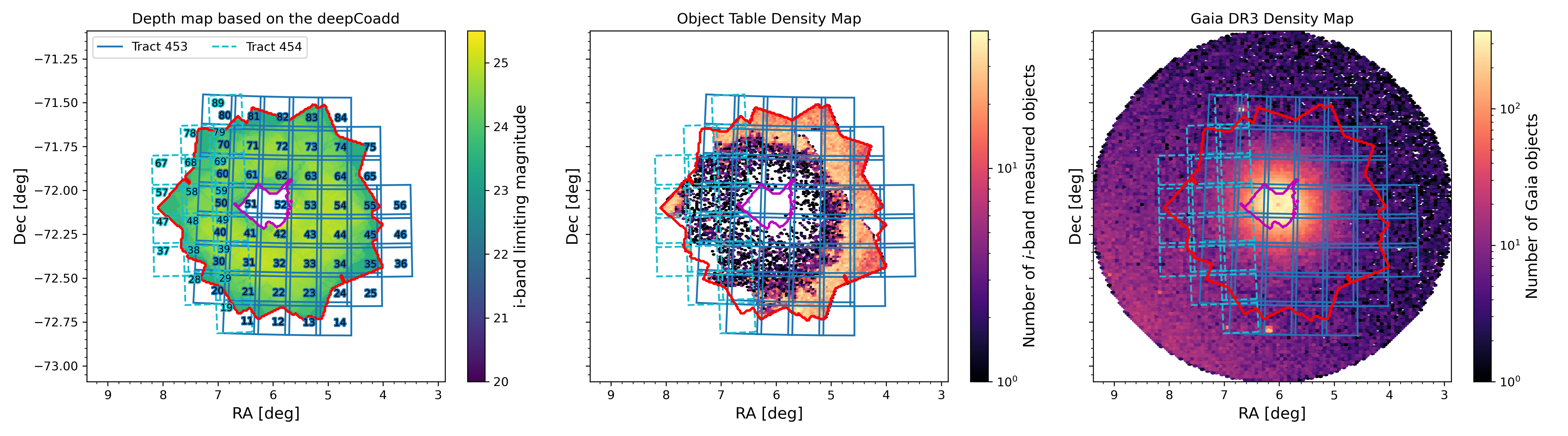}
       \caption{\textit{Left:} Same as Fig.~\ref{fig:depthcoverage}, but in the $i$ band. Patch numbers are labeled using the same outline color as their corresponding patch boundaries. \textit{Middle:} Spatial density map of all objects with measurements in the $i$ band in the 47 Tuc field, based on the \texttt{Object} table retrieved using the ADQL query described in Sec.~\ref{sec:objtab}. Unlike the limiting magnitude map estimated assuming the ideal case, it is discontinuous, with visible gaps, and there is a sharp drop in the number of detected objects near the center. This is highly likely due to crowding. In particular, the number of objects decreases further toward increasing RA, likely because crowding is exacerbated by increased contamination from SMC stars. \textit{Right:} Spatial density map of Gaia DR3 objects within 1 degree of the 47 Tuc field center. The enhanced density on the eastern side reflects increased contamination from the SMC.  
      \label{fig:spatial_mismatch}}
\end{figure*}

\subsubsection{Depth and Coverage}
Figure~\ref{fig:depthcoverage} shows the cumulative $r$-band imaging depth (i.e., coadd depth), expressed as the 5$\sigma$ limiting magnitude, for the 47 Tuc field. This map effectively represents an idealized PSF magnitude limit, assuming perfect detection and measurement. The outlines of coadd patches are overlaid, with each tract distinguished by a unique color and line style. The DP1 skymap partitions the celestial sphere into 18,938 tracts, each spanning approximately 2.8 square degrees. Each tract is further subdivided into a 10 × 10 grid of uniformly sized patches, with each patch covering roughly 0.028 square degrees \citep{RTN-095}. The bright core of 47 Tuc is saturated in the LSSTComCam commissioning data, leading to a lack of coadd coverage in that region (see the hole in Fig.~\ref{fig:depthcoverage}). The absence of coadds in that region implies that no object measurements, difference imaging, difference-image source analysis, and forced photometry analysis could be performed.

To estimate the per-band areal coverage of the coadd images of the field, we identified valid regions in the depth map and traced their boundaries using a contour-based approach. This method captures the detailed shape of the coverage, including any internal gaps or holes. The resulting contours were converted to sky coordinates and used to construct a polygonal representation of the footprint. The total area was then computed from this geometry both including and excluding the saturated central region. The outermost boundary and the central hole determined for the $r$ band are shown in Figure~\ref{fig:depthcoverage}. The outer boundary traces the full extent of the valid data, while the inner boundary delineates the saturated region excluded from the hole-corrected area measurement. The resulting sky coverage in the $r$ band is approximately 1.064~deg$^{2}$ with the central hole included, which spans 0.149~deg$^{2}$, and 0.915~deg$^{2}$ when the hole is excluded. The sizes of the central holes in the $g$, $i$, and $y$ bands are 0.155~deg$^{2}$, 0.169~deg$^{2}$, and 0.060~deg$^{2}$, respectively. The areal coverage including and excluding the saturated region in other bands is listed in Table~\ref{tab:47tuc_filters}.

\subsubsection{Impact of Crowding on Deblending}\label{sec:objtab}
In this and following sections, we explore the \texttt{Object} table to provide an overview of selected properties that are most relevant for analyzing the stellar populations in this crowded field. The \texttt{Object} table stores measurements of objects detected in the coadd images with a signal-to-noise ratio (SNR) of $>$5 in at least one band, and contains 1,296 columns \citep{Objecttable}. 

The ADQL query used in this analysis is as follows:
\begin{lstlisting}[language=SQL]
SELECT objectId, coord_ra, coord_dec, ebv, refExtendedness, patch, g_inputCount, r_inputCount, i_inputCount, y_inputCount, scisql_nanojanskyToAbMag(g_psfFlux) as g_psfmag, scisql_nanojanskyToAbMagSigma(g_psfFlux, g_psfFluxErr) as g_psfmagerr, scisql_nanojanskyToAbMag(g_cModelFlux) as g_cModelmag, scisql_nanojanskyToAbMag(r_psfFlux) as r_psfmag, scisql_nanojanskyToAbMagSigma(r_psfFlux, r_psfFluxErr) as r_psfmagerr, scisql_nanojanskyToAbMag(r_cModelFlux) as r_cModelmag, scisql_nanojanskyToAbMag(i_psfFlux) as i_psfmag, scisql_nanojanskyToAbMagSigma(i_psfFlux, i_psfFluxErr) as i_psfmagerr, scisql_nanojanskyToAbMag(i_cModelFlux) as i_cModelmag, scisql_nanojanskyToAbMag(y_psfFlux) as y_psfmag, scisql_nanojanskyToAbMagSigma(y_psfFlux, y_psfFluxErr) as y_psfmagerr, scisql_nanojanskyToAbMag(y_cModelFlux) as y_cModelmag
FROM dp1.Object
WHERE CONTAINS(POINT(`ICRS',coord_ra,coord_dec),CIRCLE(`ICRS', 6.128, -72.09, 1.0)) = 1
\end{lstlisting}
The query returned 115,493 objects.

Figure~\ref{fig:spatial_mismatch} shows the i-band 5$\sigma$ limiting magnitude map (left), the number count map of $i$-band measured objects in the \texttt{Object} table (middle), and the number count map of all Gaia DR3 objects within 1 degree from the 47 Tuc field (right). As mentioned earlier, the depth map is estimated from the properties of the images contributing to the coadds assuming the ideal detection and measurement scenario, while the density map include all entries in the \texttt{Object} table for detections in the coadd images with SNR $>$ 5 in at least one band. In contrast to the limiting magnitude map, the \texttt{Object} table density map is discontinuous, with noticeable gaps in addition to the expected central hole, and a sharp decline in the number of detected objects near the field center. The decline becomes more pronounced toward increased RA, consistent with enhanced stellar contamination from the SMC in that region, as shown in the Gaia density map. The SMC is centered at (RA, Dec) = (13.187, -72.8297) degrees, with the angular separation of 2.578 degrees from the 47~Tuc field. 

To understand the origin of the apparent lack of sources in the central and eastern parts of the field, we examine deblending outcomes for two representative tract/patch pairs: (454, 48), located on the eastern side where the resulting object density is low, and (453, 44), on the western side, which includes both dense and sparse regions. 

\begin{figure}
 \centering
      \includegraphics[width=\linewidth]{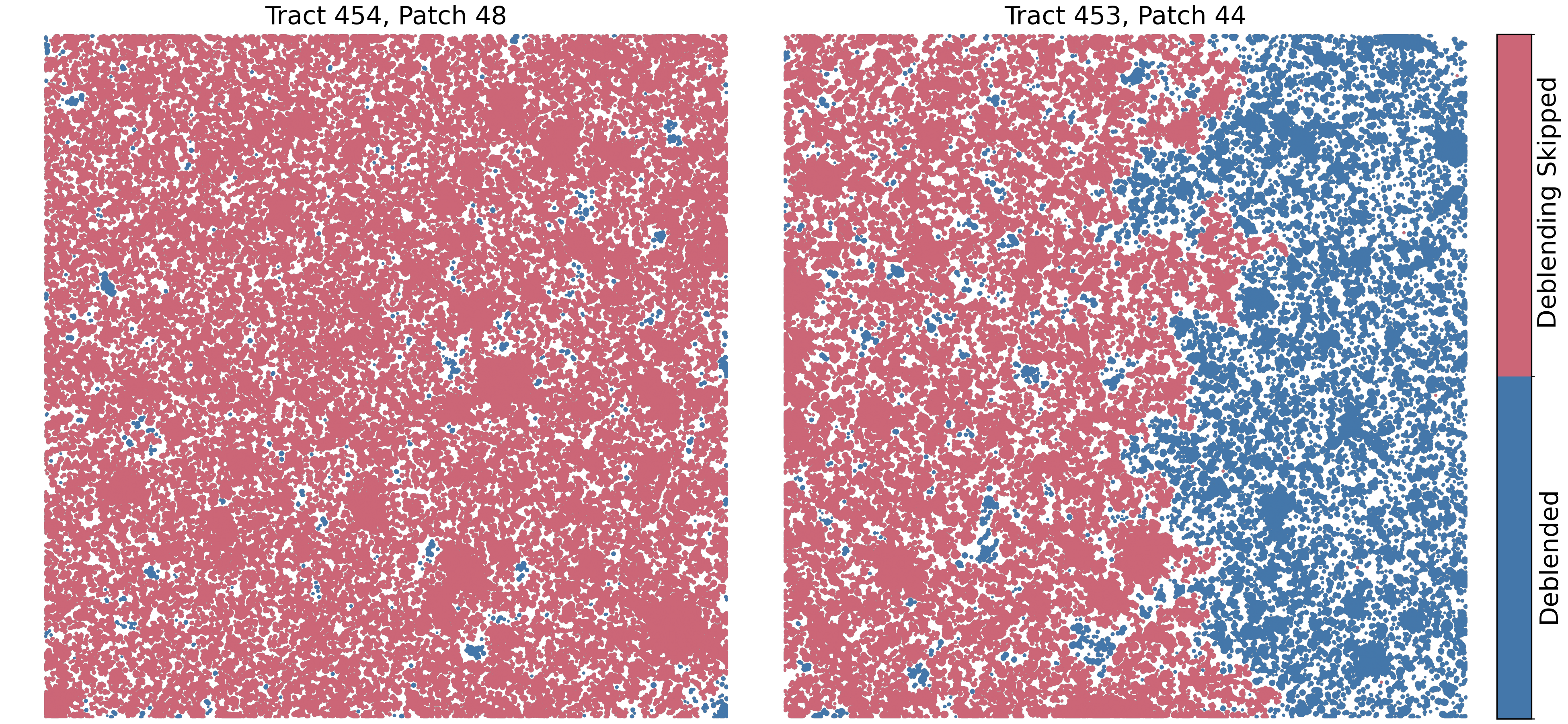}
       \caption{Comparison of deblending outcomes for two example tract/patch pairs: (454, 48), which exhibits severe crowding, and (453, 44), which contains a mix of crowded and less crowded regions. Successfully deblended sources are shown in blue, while those where deblending was skipped for various reasons, including crowding, are shown in pink. 
      \label{fig:deblend}}
\end{figure}

Figure~\ref{fig:deblend} shows the spatial distribution of sources for which the deblending process was either performed or skipped due to excessive memory consumption or runtime requirements. Deblending can be skipped for various reasons, including excessively large parent footprints, too many detected peaks, a single-peak configuration, or excessive masking. 

In the eastern region (left panel, tract: 454, patch:48), the notable drop in \texttt{Object} table source counts appears to be primarily driven by skipped deblending, likely due to additional crowding associated with the proximity to the SMC. In such regions, the source detection algorithm correctly identifies peaks, but the majority of pixels in the patch are above the detection threshold, resulting in one giant detection footprint with a few small holes. This occurs when the footprints of many overlapping sources merge into a single, large blend, and the current implementation of \texttt{scarlet} lite skips these blends due to memory constraints. This failure mode arises not simply from stellar density, but from a combination of crowding and the brightness of overlapping sources, i.e., a few bright stars can trigger the same behavior as a dense cluster of faint ones. This underscores the inherent challenges of performing photometry in crowded fields, a regime for which the LSST Science Pipelines are not specifically optimized. Improvements to \texttt{scarlet} \citep{Melchior2018} lite should allow processing to push further into these and other dense regions before the first data release. 

In contrast, in the western region (right panel, tract: 453, patch:44), deblending was largely skipped in the crowded inner area but completed successfully in the outer parts, where the intrinsic object density is lower. The transition is pretty sharp, occurring almost vertically at RA = 4.85~deg. Unfortunately, measuring an accurate, representative crowding level at which the pipelines begin to struggle in DP1 is non-trivial due to complex spatial coverage variations and incompleteness that cannot be characterized without artificial star tests, which is beyond the scope of this paper.

%====== Figure

%====== Figure
\begin{figure}
 \centering
      \includegraphics[width=\linewidth]{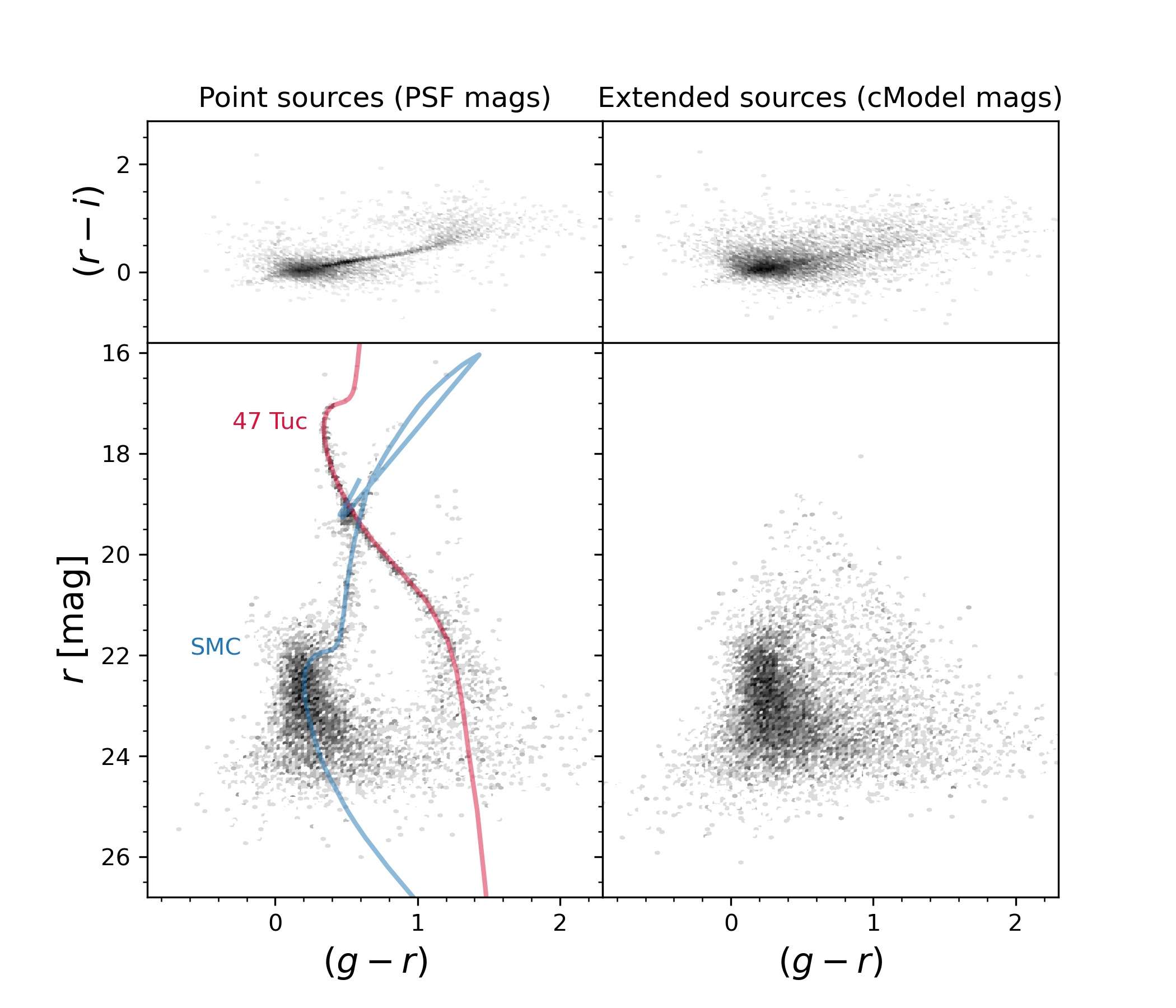}
       \caption{Top panels display $(r - i)$ vs. $(g - r)$ color-color diagrams and bottom panels show $r$ vs. $(g - r)$ color-magnitude diagrams for objects classified by morphology. The left column shows likely stars (\texttt{refExtendedness} $=$ 0) using PSF magnitudes, and the right column shows likely galaxies (\texttt{refExtendedness} $=$ 1) using cModel magnitudes. The red isochrone represents the old stellar population of 47 Tuc, with an age of 12.4~Gyr, [M/H] = -0.5, and distance of 4.66~kpc \citep{Simunovic2023}. The blue isochrone traces a diffuse stellar populations from the SMC, with an age of 8~Gyr, [M/H] = -1, and distance of 62~kpc. Both isochrones are generated using PARSEC version 1.2S \citep{Bressan2012, Tang2014, Chen2015}.
      \label{fig:CCDsCMDs}}
\end{figure}

%====== Figure
\begin{figure*}
 \centering
      \includegraphics[width=\linewidth]{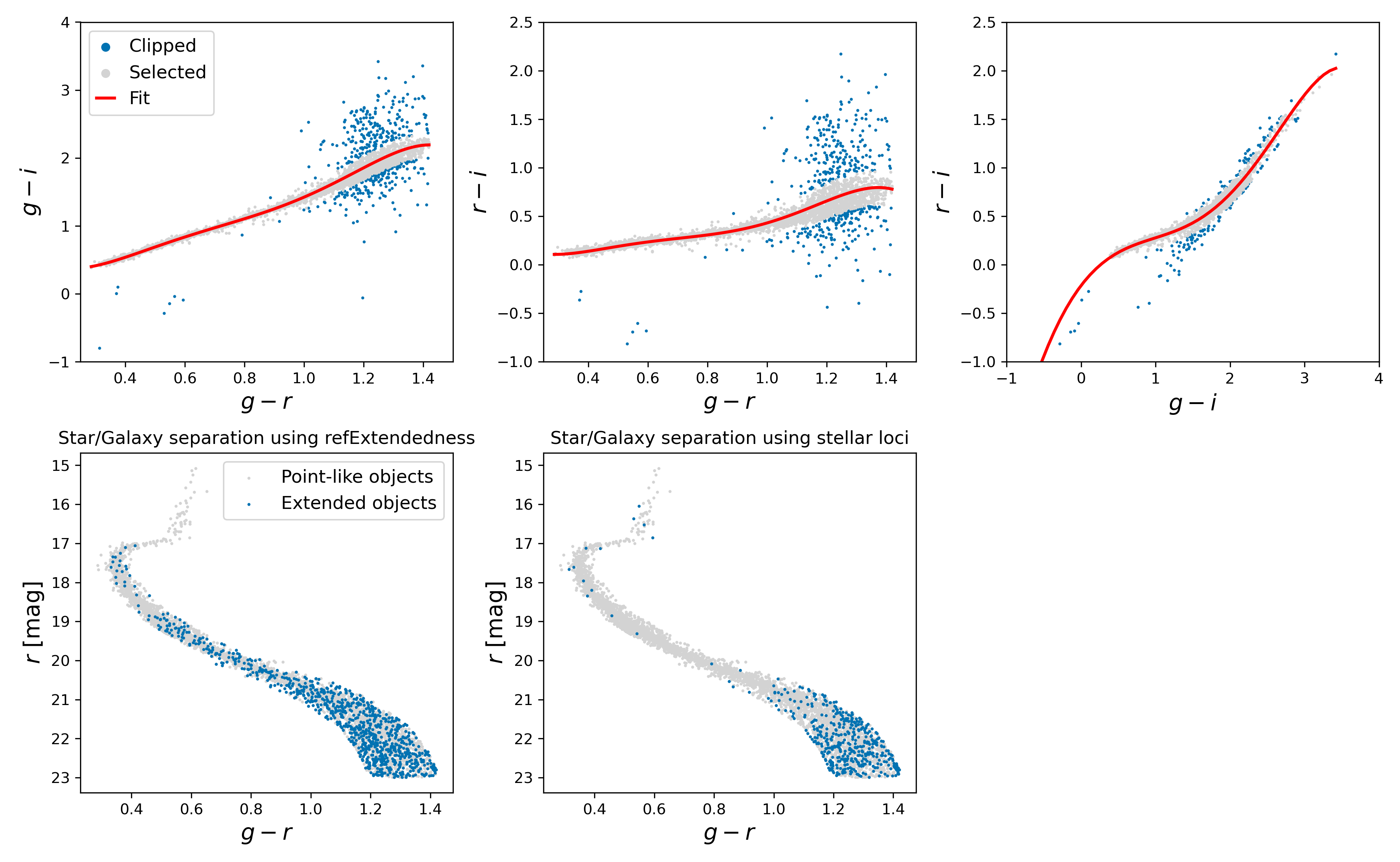}
       \caption{Star/Galaxy separation in color--color spaces. Top panels display the three color-color diagrams: ($g - r$, $g - i$), ($g - r$, $r - i$), and ($g - i$, $r - i$). In each panel, a fifth-degree polynomial was fit to the observed stellar locus (red line), using a subset of stars preselected based on isochrone proximity and moderate color cuts to reduce extreme outliers. Objects with residuals exceeding 1$\sigma$ from the fitted locus (blue dots) are classified as likely background galaxies, while those within 1$\sigma$ (light gray dots) are retained as probable stars. Bottom panels compare the star/galaxy separation using the \texttt{refExtendedness} parameter and our multi-dimensional sigma-clipping method. The \texttt{refExtendedness}-based approach tends to misclassify bright stars as extended objects, whereas our method primarily rejects faint contaminants consistent with background galaxies.    
      \label{fig:stargalaxy}}
\end{figure*}

\subsubsection{Star-Galaxy Separation}\label{sec:stargalaxy}
Unresolved and resolved objects often occupy the same part of the sky but represent fundamentally different physical systems. Misclassifying galaxies as stars (or vice versa) leads to systematic errors in object catalogs and thus biased measurements of stellar populations, galaxy properties, and large-scale structure. In crowded fields like 47 Tuc, where stars and compact galaxies overlap in morphology and color especially at fainter magnitude, robust classification becomes especially critical for downstream science.

The LSST Science Pipelines report an extendedness parameter, which is based on the ratio of composite model \citep[cModel;][]{Abazajian2004, Bosch2018} and PSF fluxes; point sources should have roughly equal fluxes from these two measures, while extended sources will have more flux in the model measurement than the PSF. Objects are classified as ``not extended'' (\texttt{extendedness}=0) if the ratio between their PSF and cModel fluxes satisfies $f_{\rm psf} > 0.985 \times f_{\rm cmodel}$. In this work, we use the \texttt{refExtendedness} parameter, which for each object corresponds to the extendedness measured in the ``reference band'' (selected as described in Sec. 3.4 of \citealt{Bosch2018}). 

Figure~\ref{fig:CCDsCMDs} shows $(r - i)$ vs. $(g - r)$ color-color diagrams (top panel) and $r$ vs. $(g - r)$ color-magnitude diagrams (bottom panel), separated by extendedness classification. The left column displays likely stars (\texttt{refExtendedness} $=$ 0) using PSF magnitudes, while the right column shows likely galaxies (\texttt{refExtendedness} $=$ 1) using cModel magnitudes. In the color-magnitude diagram of likely stars, three components are evident: 47 Tuc, the SMC, and the Milky Way. We overlay PARSEC v1.2S isochrones on the stellar locus of 47 Tuc and the SMC \citep{Bressan2012, Tang2014, Chen2015}. The red isochrone represents the old stellar population of 47 Tuc, with an age of 12.4~Gyr, [M/H] = -0.5, and distance of 4.66~kpc \citep{Simunovic2023}. The blue isochrone traces diffuse stellar populations from the SMC, with an age of 8~Gyr, [M/H] = -1, and distance of 62~kpc. Traces of both the 47 Tuc and SMC sequences remain visible in the color-magnitude diagram of extended objects, indicating that star-galaxy separation using the \texttt{refExtendedness} parameter becomes unreliable at fainter magnitudes in crowded fields such as 47 Tuc. 

Color-color diagrams are a powerful tool for separating stars from unresolved background galaxies, as stars occupy well-defined, continuous loci in multiple color-color spaces, while galaxies and other contaminants typically deviate from these loci. This distinction becomes especially useful in deep imaging where morphological classification becomes unreliable at faint magnitudes. To leverage this, we performed polynomial fits to the stellar locus in three color--color spaces: $(g - r,\ g - i)$, $(g - i,\ r - i)$, and $(g - i,\ r - i)$. 

Figure~\ref{fig:stargalaxy} illustrates the process and compares the multiple color-color diagram-based star--galaxy separation against that based on the \texttt{refExtendedness} parameter. The $y$ band was excluded due to its limited depth. For each pair, we selected likely 47~Tuc stars using the isochrone presented in Figure~\ref{fig:CCDsCMDs} (see Section~\ref{sec:isoselection}) and moderate color ranges to exclude extreme outliers and poorly measured objects. A fifth-degree polynomial was fit to each observed stellar locus. We then computed the residuals of all objects from the fitted curve. Objects with residuals exceeding 5$\sigma$ were identified in each color--color space, allowing us to exclude sources that lie significantly off the stellar locus in all three spaces. The resulting polynomial fits are:
\begin{align}
(g - i) &= -6.835\,(g - r)^5 + 26.964\,(g - r)^4 - 39.679\,(g - r)^3 \nonumber \\
        &\quad + 27.316\,(g - r)^2 - 7.381\,(g - r) + 1.039 \\
(r - i) &= -5.793\,v^5 + 22.556\,(g - r)^4 - 32.630\,(g - r)^3 \nonumber \\
        &\quad + 22.024\,(g - r)^2 - 6.521\,(g - r) + 0.794 \\
(r - i) &= -0.0137\,(g - i)^5 + 0.0333\,(g - i)^4 + 0.244\,(g - i)^3 \nonumber \\
        &\quad - 0.781\,(g - i)^2 + 1.009\,(g - i) - 0.215.
\end{align}

In the bottom panels of Figure~\ref{fig:stargalaxy}, we compare the star/galaxy separation using the pipeline-measured \texttt{refExtendedness} parameter and our multi-dimensional sigma-clipping method. The \texttt{refExtendedness}-based classification misidentifies highly-likely, relatively bright stars along the isochrone as extended sources, leaving an imprint of the 47~Tuc stellar locus in the extended-object color-magnitude diagram, as shown in the bottom right panel of Figure~\ref{fig:CCDsCMDs}. In contrast, our approach more reliably excludes faint objects with colors inconsistent with the stellar locus, likely corresponding to background galaxies.

\section{Analysis}\label{sec:analysis}
In this section, we describe the membership determination procedure, including the improved star-galaxy separation described in Section~\ref{sec:stargalaxy}, and examine the DP1 photometry for the resulting sample of cluster member stars. We also investigate the detection of variable stars in the 47 Tuc field.

\subsection{Selecting 47 Tuc Member Stars}
\subsubsection{Isochrone-based Selection}\label{sec:isoselection}
As shown in Figure~\ref{fig:CCDsCMDs}, the 47 Tuc sequence is well defined and closely follows the model isochrone in the color-magnitude diagram. We therefore identified likely member stars by measuring their proximity to a reference isochrone in color-magnitude diagram space. For each observed star, we computed the minimum distance to the isochrone, normalized by uncertainties in both color and magnitude. Tighter color tolerances (0.05~mag) were applied to bright stars ($r <$ 20~mag) to reflect their higher photometric precision, while fainter stars used a broader threshold of 0.1~mag. 

We converted these distances into a likelihood-like score using a standard normal distribution. The closer the star is to the fiducial isochrone, the higher the matched scores. To identify probable member stars, we selected those exceeding threshold of 0.2. This method effectively isolated high-confidence candidates in the color-magnitude diagram space, leaving 4,276 candidate stars. However, in regions of the color-magnitude diagram where 47 Tuc main sequence overlaps with the SMC red clump, this approach alone is insufficient. To mitigate contamination in that regime, we applied an additional cut based on Gaia DR3 proper motion measurements \citep{GaiaDR32023}. 

\subsubsection{Gaia Proper Motion-based Selection}
We retrieved Gaia DR3 proper motions and photometry using the same center and radius as the \texttt{Object} table query, excluding sources with extreme proper motions or \texttt{astrometric\_excess\_noise} $\geq$ 0.2, yielding 102,902 objects. Given Gaia's limiting magnitude of $G \approx 21$, the goal in this step is not to identify stars consistent with the 47~Tuc proper motion, but to exclude clear outliers in the proper motion space from the isochrone-selected sample. 

\begin{figure}
 \centering
      \includegraphics[width=\linewidth]{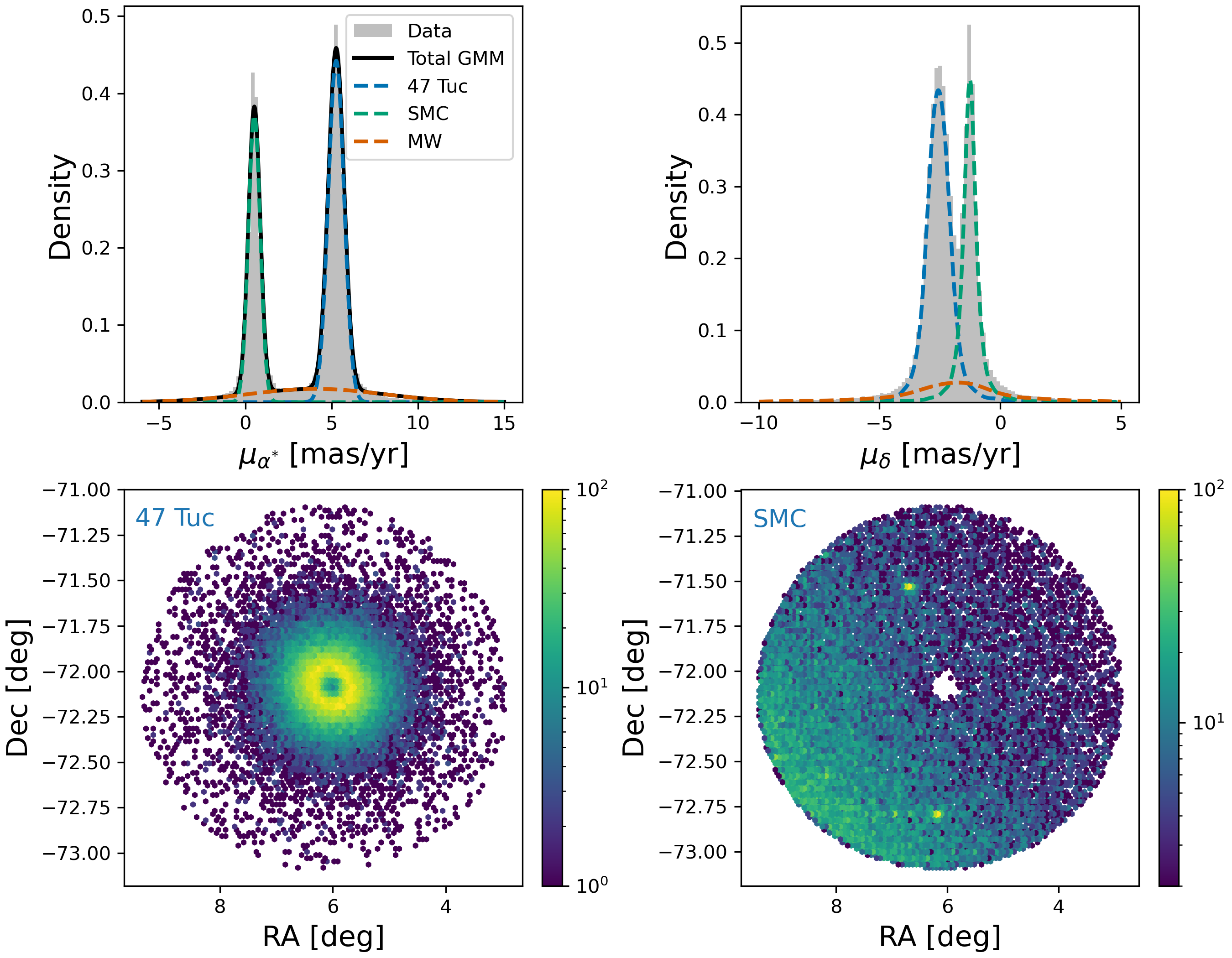}
       \caption{\textit{Top:} Normalized histograms of $\mu_{\alpha^{*}}$ and $\mu_{\delta}$ with the fitted GMM overlaid, scaled by their relative weights. The total GMM fit is shown in black, with individual kinematic components representing the Milky Way (orange), 47 Tuc (blue), and the SMC (green). Although the model was fit only in $\mu_{\alpha^{*}}$, the components show clear separation in $\mu_{\delta}$ as well. \textit{Bottom:} Spatial distributions of Gaia DR3 objects with proper motions consistent with the 47~Tuc component (left) and the SMC component (right). The 47 Tuc map shows a strong central concentration, while the SMC map exhibits a density gradient increasing toward the SMC center.   
      \label{fig:gaiaPM}}
\end{figure}

Figure~\ref{fig:gaiaPM} demonstrates the separation of stellar populations in proper motion space based on a Gaussian Mixture Model (GMM) classification. We applied the GMM to the distribution of proper motions in RA ($\mu_{\alpha^{*}}$) to identify kinematic components in the 47 Tuc field using \texttt{sklearn.mixture.GaussianMixture}. The model was fit using only the $\mu_{\alpha^{*}}$ values, assuming three Gaussian components to represent the Milky Way contaminants, the 47 Tuc cluster, and the background SMC populations. The resulting classification was then used to label stars and examine the corresponding distributions in DEC proper motion ($\mu_{\delta}$). Although the GMM was trained solely on $\mu_{\alpha^{*}}$, the $\mu_{\delta}$ distributions for each component remain well distinct, further validating the classification. In the top panels, we present these distributions by overplotting scaled Gaussian kernel density estimates of $\mu_{\alpha^{*}}$ and $\mu_{\delta}$ for each component identified by the $\mu_{\alpha^{*}}$-based GMM, weighted by the relative number of stars assigned to each group. The bottom panels show spatial density maps of Gaia DR3 objects with proper motions consistent with the 47~Tuc component (left) and the SMC component (right). A total of 308 stars were excluded from the isochrone-selected sample during this proper motion filtering step.

%====== Figure

%====== Figure
\begin{figure*}
 \centering
      \includegraphics[width=\linewidth]{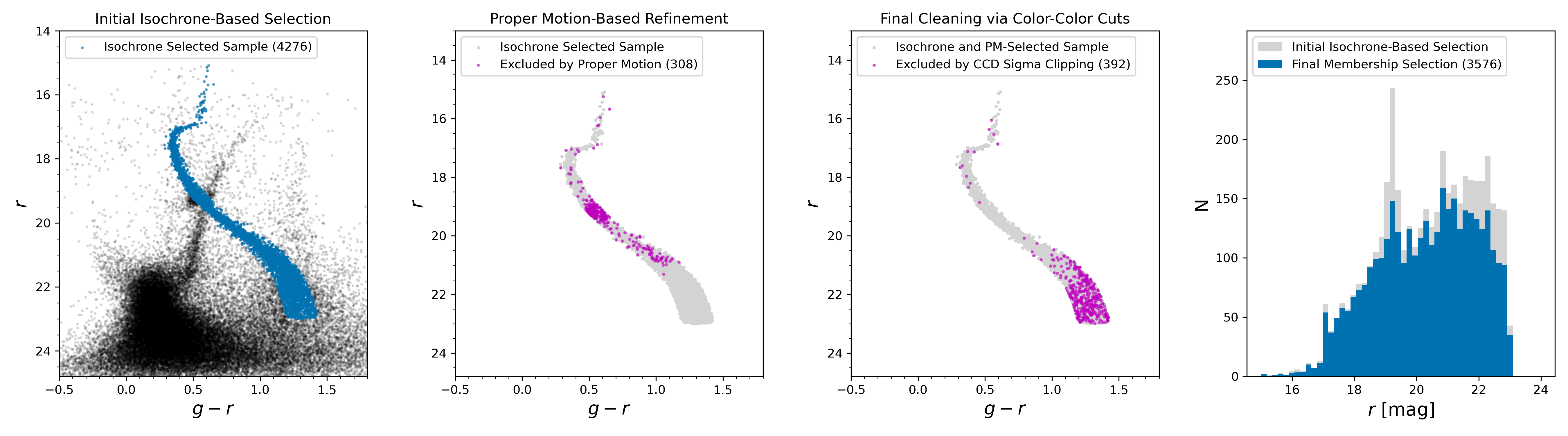}
       \caption{Summary of the sequential member selection process. The first three panels show color-magnitude diagrams after each selection step: (1) isochrone-based pre-selection, (2) proper motion refinement, and (3) color-color diagram-based filtering. The fourth panel shows the $r$-band magnitude distributions of the initial isochrone-selected sample (gray) and the final sample of 3,576 (blue), highlighting the removal of bright SMC contaminants via Gaia proper motion filtering and faint background galaxies through color-color diagram-based selection. A catalog of unique object IDs for the identified cluster member stars is available at https://zenodo.org/records/15794218. 
      \label{fig:member_47tuc}}
\end{figure*}

\subsubsection{Selection in Color-color Diagrams}
To further refine the membership sample, we applied the same sigma-clipping method in the three color-color spaces, using the polynomial fits determined in Section~\ref{sec:stargalaxy}. This multi-dimensional color-color filtering removed a total of 392 objects that deviate significantly from the stellar locus, helping eliminate any remaining contaminants with inconsistent colors. 

Figure~\ref{fig:member_47tuc} presents a summary of the sequential member selection process. The first three panels show color-magnitude diagrams illustrating the impact of each filtering step, from left to right: isochrone-based pre-selection, proper motion refinement, and final color-color diagram filtering. Each panel highlights the progressive narrowing of the stellar locus and removal of contaminants. The fourth panel compares the $r$-band magnitude distribution of the initial isochrone-selected sample and the final sample after all filtering steps. The proper motion refinement effectively removed significant contamination from the SMC, particularly around its red clump, as well as brighter stars down to Gaia’s depth limit. In contrast, the color-color diagram filtering step primarily excluded faint objects—likely background galaxies—that could not be identified through Gaia proper motion due to their low brightness. The final sample consists of 3,576 member stars. The list of the selected 47 Tuc member stars is available in a machine-readable format\footnote{https://zenodo.org/records/15794218}. The catalog provides only the unique object IDs, that can be crossed matched to the DP1 data products via the Rubin Science Platform \citep[RSP;][]{LSE-319,DMTN-212}.

\subsection{Exploring Photometry of 47 Tuc}
\subsubsection{Saturation Limit}
Saturation limits define the upper reliability threshold of the data and are essential for quality control, completeness assessment, and robust scientific interpretation. Saturated stars yield inaccurate or unusable flux measurements, and without a clear understanding of the saturation limit, bright sources can be mischaracterized, biasing downstream analyses. Even with identical exposure times (as in the DP1 case) and identical spectral types, the saturation limit of a star can vary with seeing, total filter throughput, background level, and other observational conditions. Moreover, as discussed in Section~\ref{sec:objtab}, the LSST Science Pipelines in DP1 were not particularly optimized for crowded fields, unfortunately leading to skipping the deblending process for sources even under modest crowding. Consequently, the brightest star with reliable measurements in the 47~Tuc field does not necessarily reflect the true saturation limit. 

The brightest 47~Tuc member star (objectId: 579576081061783277) is on the red giant branch phase and has magnitudes of $15.6961\pm0.0007$, $15.0811\pm0.0003$, $14.9405\pm0.0006$, and $14.7374\pm0.0007$ in the $g$, $r$, $i$, and $y$ bands, respectively. The star has a Gaia counterpart, and its transformed photometry from the Gaia $G$ and $BP-RP$ to the LSSTComCam system \citep{RTN-099} agrees well within a few hundredths of a magnitude in the $g$, $r$, and $y$, where no visit images have saturated pixels. In the $i$ band, however, the transformed magnitude differs by $\sim$0.06~mag, coinciding with mild saturation in the central pixels of a few visits. This offset is likely due to the fact that the $i$ band has the best seeing in the field (see Figure~\ref{fig:psf_limitmags} and Table~\ref{tab:47tuc_filters}), leading to saturation in the core, and the RGB star peaks in flux near that band. The second brightest $i$-band magnitude without saturated pixels is 15.1942$\pm$0.0002. In short, the unsaturated bright limits reported in the \texttt{Object} table for 47 Tuc are approximately 15.70, 15.08, 15.19, and 14.74~mag in the $g$, $r$, $i$, and $y$ bands, respectively. 

As a check on the onset of nonlinearity in the photometry at the bright end, we examined the color-magnitude diagrams of the low Galactic latitude field Rubin\_SV\_095\_-25.  The sequence of numerous red disk turnoff stars in the color-magnitude diagrams of this field departs from its expected vertical shape at the bright end at magnitudes of $griy\sim$15.9, 16.2, 15.8, 15.7, an indication of the point at which non-linearity begins to affect the photometry of bright stars.  These limits are up to a magnitude fainter than seen in 47 Tuc, such that care should be taken with photometry brighter than $\sim$16th magnitude.

%====== Figure
\begin{figure*}[t]
 \centering
      \includegraphics[width=\linewidth]{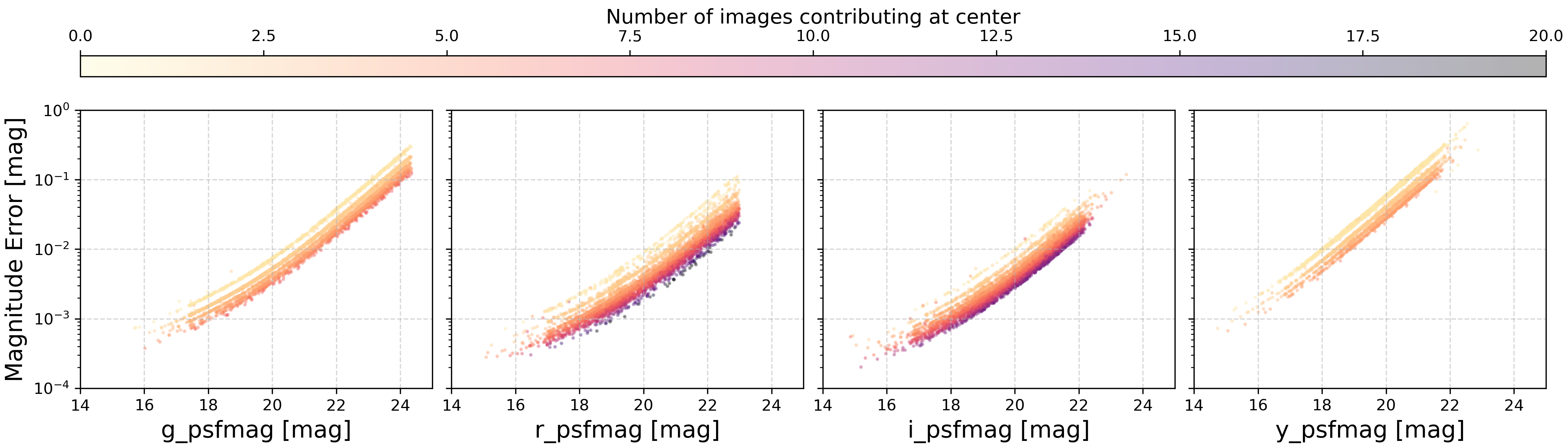}
       \caption{Photometric uncertainty as a function of PSF magnitude for selected member stars in each band. The discrete sequences arise from variations in image depth. Each panel is color-coded by the number of input images contributing to the photometry at each object's position. Objects with more input images show smaller uncertainties. The correspondence between sequence structure and input image count confirms that image depth is the dominant factor responsible for theses patterns. 
      \label{fig:photometry}}
\end{figure*}

\subsubsection{Photometric Uncertainty}
Figure~\ref{fig:photometry} demonstrates the photometric uncertainty of the selected member stars in each band. While the shallowest $y$-band photometry show relatively large uncertainty, the other three bands achieve 1\% level precision roughly down to 21.5, 22, 21.5~mag in the $g$, $r$, $i$ bands, respectively. The multiple quantized sequences seen in each band reflect variation in image depth across the field. Photometry from deeper exposures (i.e., more input images) yields smaller uncertainties. To illustrate this, each panel is color-coded by the number of input images contributing to the measurement at each object's position. The distinct sequences correspond to different numbers of input images, confirming that depth is the dominant factor shaping the structure of these sequences.

\subsubsection{Analysis of Photometric Scatter}
\begin{figure}
 \centering
      \includegraphics[width=\linewidth]{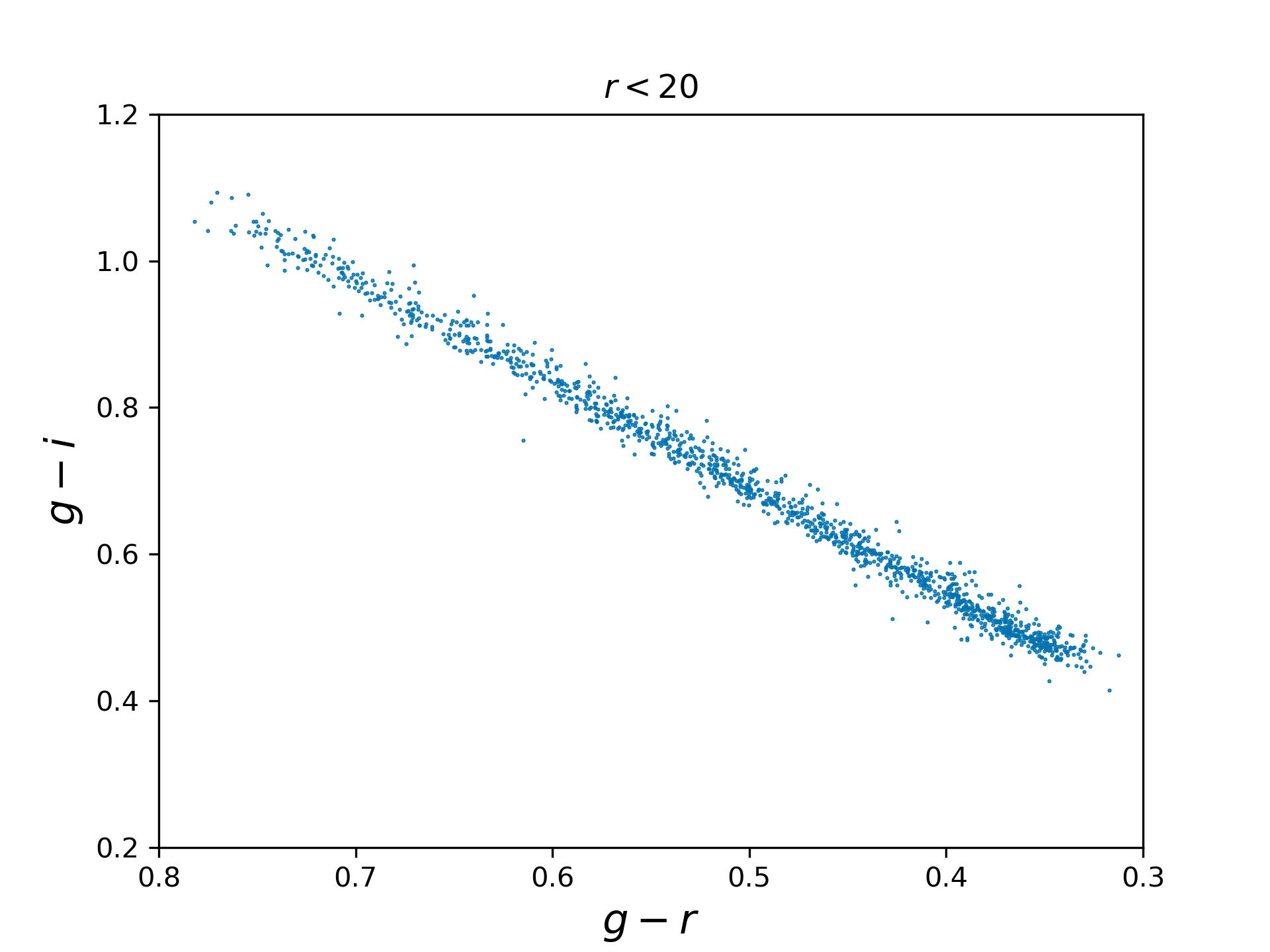}
       \caption{Color-color diagram for bright stars in 47 Tuc. The tight linear sequence for point-like sources with $r<20$ is expected for a simple stellar population.
      \label{fig:colorcolor}}
\end{figure}

The ``simple" stellar population of 47 Tuc affords the opportunity to diagnose possible sources of scatter in the DP1 photometry.  Figure~\ref{fig:colorcolor} shows the $g - i$ vs.\ $g - r$ color-color diagram for bright stars ($r<20$), which features a tight linear sequence.  We considered five possible sources to explain the scatter about this sequence: 1) PSF-fitting errors stemming from known Poisson and other observational noise sources; 2) crowding errors \citep[e.g.,][]{Olsen2003}; 3) differential extinction by dust \citep[e.g.,][]{Pancino2024}; 4) multiple stellar populations, as observed by \citet{Milone2012} and \citet{Marino_2023}; and 5) unknown sources of observational systematic error.

\begin{figure}
 \centering
      \includegraphics[width=\linewidth]{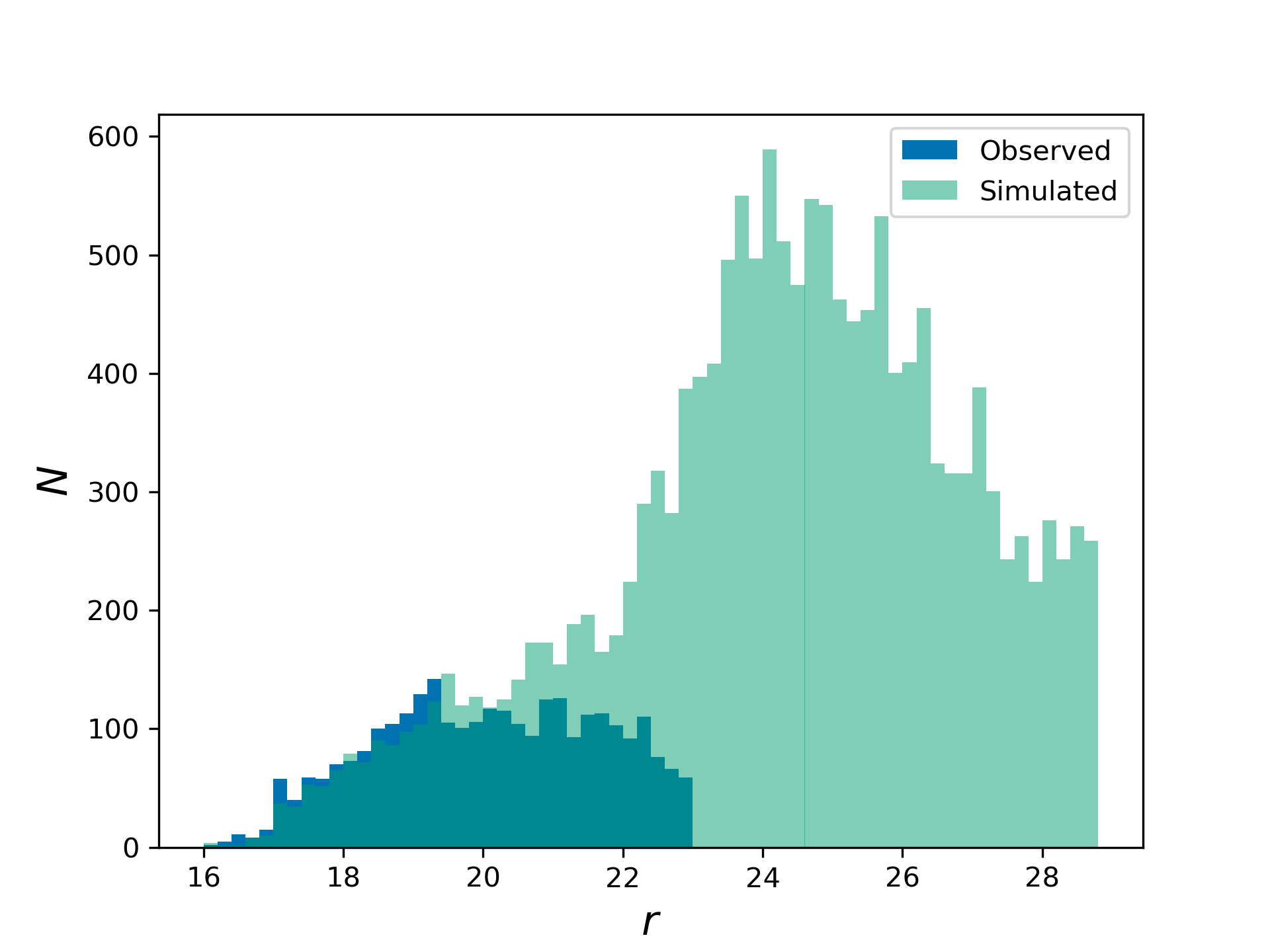}
       \caption{Observed and simulated $r$-band luminosity function for 47 Tuc.  The simulated LF derives from a PARSEC model with age 12.4 Gyr and [M/H]=$-0.5$, and has been normalized to the observed number of bright stars ($r<20$).
      \label{fig:LF}}
\end{figure}

For this analysis, we used the Padova Observatory CMD 3.8 web interface (https://stev.oapd.inaf.it/cmd) to generate a simulated stellar population with age 12.4~Gyr, [M/H]=$-0.5$ \citep{Simunovic2023}, and total mass of 10$^4$~$M_\odot$. The stellar population was drawn from isochrones based on PARSEC release v1.2S+COLIBRI S\_37+S\_35+PR16 \citep{Bressan2012,Chen2014,Tang2014,Chen2015,Rosenfield2016,Marigo2017,Pastorelli2019,Pastorelli2020} and uses the LSST throughputs Release 1.9 (Sept 2023) to compute simulated magnitudes in the LSST passbands.  Figure~\ref{fig:LF} shows the $r$-band luminosity function of the simulated population compared to the observations, where we have normalized the simulation to match the observed number of stars with $r<20$ and adopted a distance modulus $\mu_0=13.25$ and $A_V=0.1$ mag \citep{Simunovic2023}.

Our approach was to add the first three sources of photometric scatter described above (PSF-fitting errors, crowding errors, and differential extinction) to the simulated stellar population, fit a line to the $g - i$ vs.~$g - r$ relation, and measure the scatter about this relation in $g-i$ to compare with the observations.  To add the PSF-fitting errors, for each simulated star we picked a random star with the $r$-band magnitude within 0.1~mag from the observed catalog, looked up its measured photometric error, and picked a random value from a Gaussian distribution with spread $\sigma$ equal to the observed error value.

To add errors due to crowding, we used the equations developed by \citet{Olsen2003} to compute the crowding errors in the $g, r,$ and $i$ bands for each simulated star.  These equations require the full simulated input luminosity functions in the three bands and the median image quality measurements from Table~\ref{tab:47tuc_filters}.  With the crowding error scale computed from these equations as a function of magnitude, we assigned a random error to each simulated star from a Gaussian distribution of the appropriate dispersion.  We accounted for the correlated errors in crowding between bands in this calculation, such that the crowding contribution to the scatter in $g - i$ is slightly lower than that expected from the crowding error in individual bands.

To add scatter from differential reddening, we again drew random values from a Gaussian distribution in $A_V$ centered on $A_{V,0}=0.1$ mag, where we tested various values for the width $\sigma_{A_V}$ for this distribution, but truncated the distribution at 0 to avoid negative extinction values.  We used $A_\lambda/A_V$ = 1.18, 0.87, and 0.66 for $\lambda=g, r,$ and $i$, respectively, based on the \cite{Cardelli1989} extinction law with $R_V=3.1$ and the LSST filter transmission curves \citep{rubin_sim}.

\begin{figure}
 \centering             
 \includegraphics[width=\linewidth]{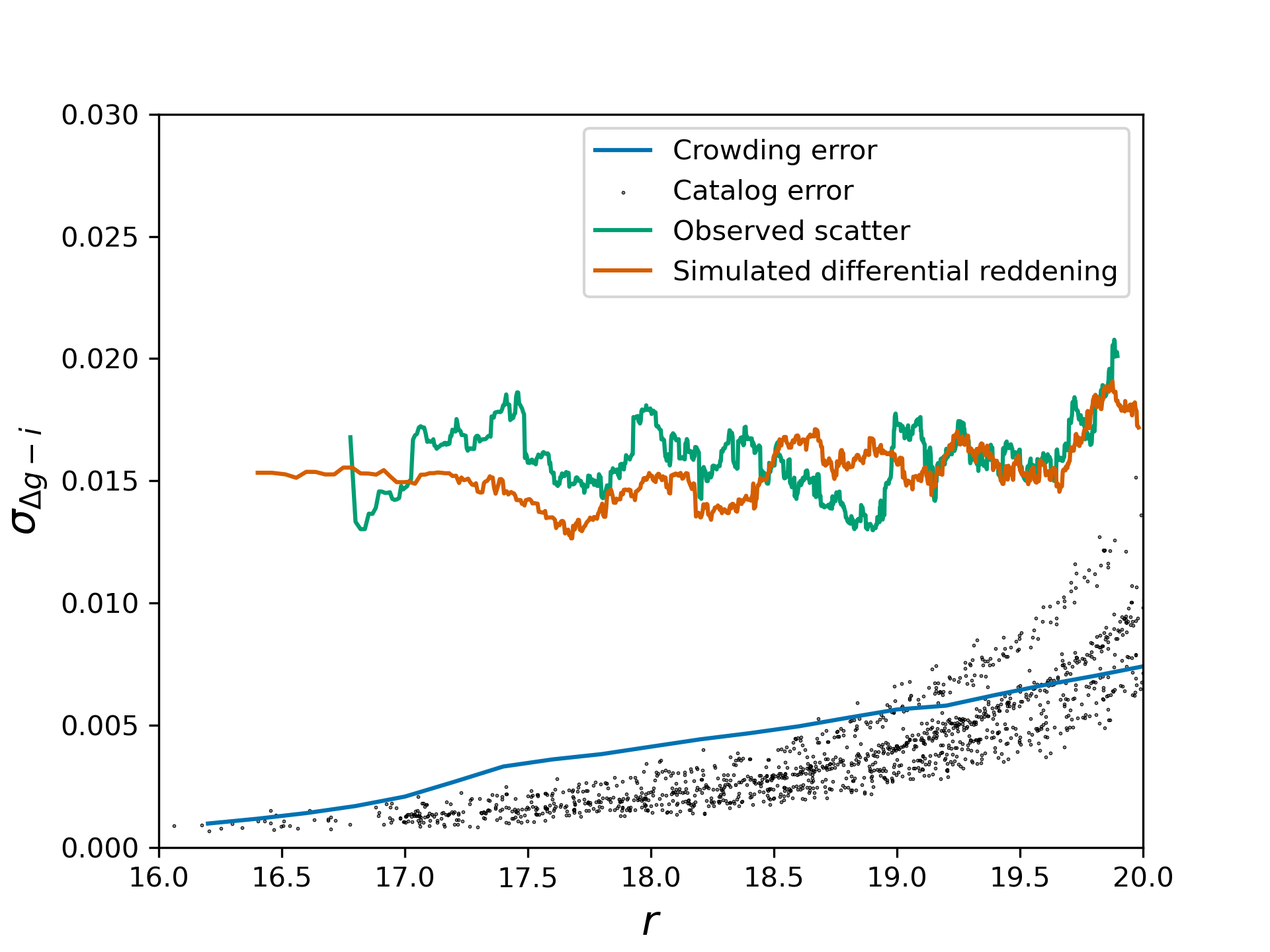}
       \caption{Possible sources of scatter in color in 47 Tuc photometry.  The observed scatter in $g - i$ color (green line) compared to a linear fit to the color-color diagram shown in Figure~\ref{fig:colorcolor} is shown along with three possible sources of the scatter: 1) the photometric errors reported for the catalog objects (black points), 2) the predicted error due to crowding (blue line), calculated as described in the text, and 3) the simulated effect of differential reddening with scale $\sigma_{A_V}=0.03$ mag (orange line).  
      \label{fig:scattercomp}}
\end{figure}

Figure~\ref{fig:scattercomp} shows the relative contributions to the scatter in $g - i$ of the simulated population about the linear fit in the $g - i$ vs.~$g - r$ plane, compared to the observed scatter.  As seen in the figure, PSF-fitting errors and crowding errors are insufficient to explain the observed scatter in $g - i$ for stars brighter than $r<20$.  Simulated differential extinction with $\sigma_{A_V}=0.03$ mag is, however, able to explain the observed scatter.  We note that this scale of differential extinction is a factor of two lower than the $\sigma_{A_V}=0.06$ mag measured by \citet{Pancino2024}, based on archival wide-field photometry.

\begin{figure}
 \centering
      \includegraphics[width=\linewidth]{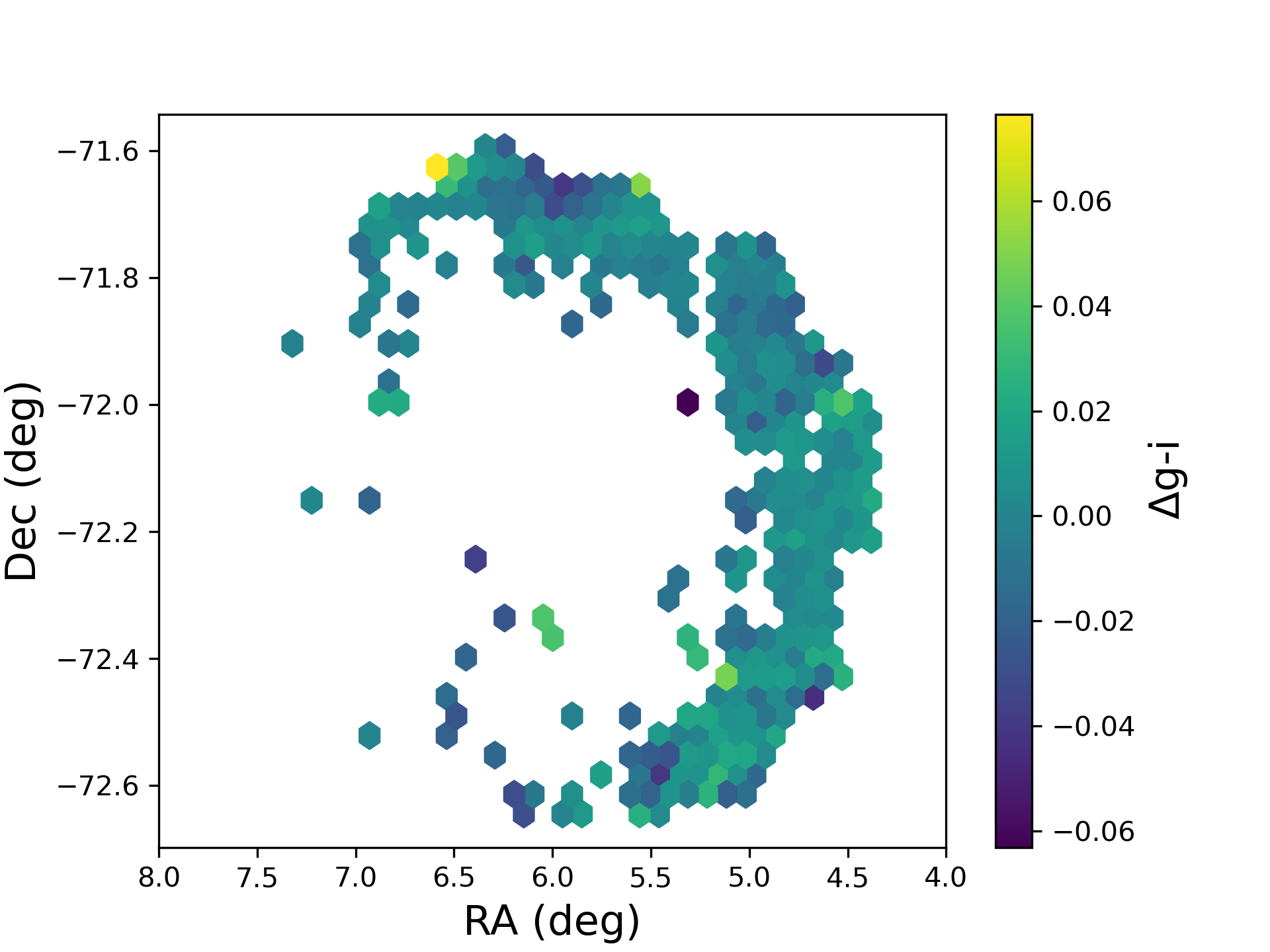}
       \caption{Map of mean offsets in color.  The map shows the mean offset in $g - i$ compared to a linear fit to the color-color diagram shown in Figure~\ref{fig:colorcolor}.  The map shows that the scatter in color predominantly stems from spatially coherent variations at scales of $\lesssim 0.1$ deg. 
      \label{fig:errormap}}
\end{figure}

We do not claim to have measured differential extinction in 47 Tuc based on modeling the photometric scatter in DP1, but we do conclude that in the areas covered, differential extinction can not be as large as seen in \citet{Pancino2024}.  Those authors cited camera-specific systematic errors as limiting their measurement, and we cannot rule out such errors in DP1.  Figure~\ref{fig:errormap} shows the observed median values of the offsets $\Delta(g - i)$ about the line fit to the $g - i$ vs.~$g - r$ plane.  The figure shows spatially coherent variations in $\Delta(g - i)$ at scales of $\lesssim 0.1$ deg, which matches the scale of the patches in the DP1 reduction.  There is not, however, obvious correlation with patch boundaries; more data will be needed to understand the source of photometric scatter in 47 Tuc. 

Multiple stellar populations are also a candidate to explain the scatter that we see.  They are known to have radial dependence \citep[e.g.,][]{Milone2012}, but would not be expected to show the azimuthal variations that we see in Figure~\ref{fig:errormap}.  With more data from LSSTCam observations of 47 Tuc, we should have an excellent opportunity to understand the source of photometric scatter in 47 Tuc, including the presence of differential extinction and multiple stellar populations.

\subsection{Variables in the 47 Tuc Field}
As an assessment of the pipeline performance and DP1 data products, we crossmatched the 47~Tuc field's \texttt{Object} and \texttt{DiaObject} \citep{DiaObject} tables to a catalog of known variables from \cite{Weldrake2004}. The original catalog contains 100 variables stars across an approximately 1~deg$^{2}$ field centered on the cluster, which is presented in their paper in Table~2. Their Tables~3--5 then detail the fitted period and amplitude of of each variable, broken out by variable type: RR Lyrae (RRLyr), Eclipsing Binaries (EcBs) and others, mostly long period variables (LPVs). Separately, a table was published to SIMBAD containing additional photometry, as well as an identifier column and links to other references for each variable. We consolidated the tables, ignoring rows with only limits or approximations of the period, amplitude or photometry, leaving 72 variables remaining. The 72 variables left include 19 EcBs, 8 detached EcBs, 11 LPVs, and 36 RRLyr. All RRLyr recovered likely belong to the SMC according to distance moduli derived by \cite{Weldrake2004}.

We then used the Large Survey Database \citep[LSDB;][]{Caplar2025} to crossmatch the 72 variable catalog to both the \texttt{Object} table and \texttt{DiaObject} table narrowed down to contain only those located in the 47 Tuc field. We find 5 matches to the \texttt{Object} table and 62 matches with the \texttt{DiaObject} table. Three variables, all RR Lyrae (ID: V23, V73, and V88), are found in both tables. The two objects not found in the \texttt{DiaObject} table correspond to one EcB and one detached EcB. As discussed in Section~\ref{sec:objtab}, the skipped debelending on the denser areas closer to the cluster core results in fewer detected objects toward the center, as shown in the middle panel of Figure~\ref{fig:depthcoverage}. In contrast, more ``DiaObjects" are detected closer to the cluster core because the difference images are less crowded after template subtraction. For our variable star crossmatching, we therefore use the \texttt{DiaObject} table, particularly because we expect there to be residual flux in the difference images due to the variability of our targets.

\begin{figure}
 \centering
      \includegraphics[width=\linewidth]{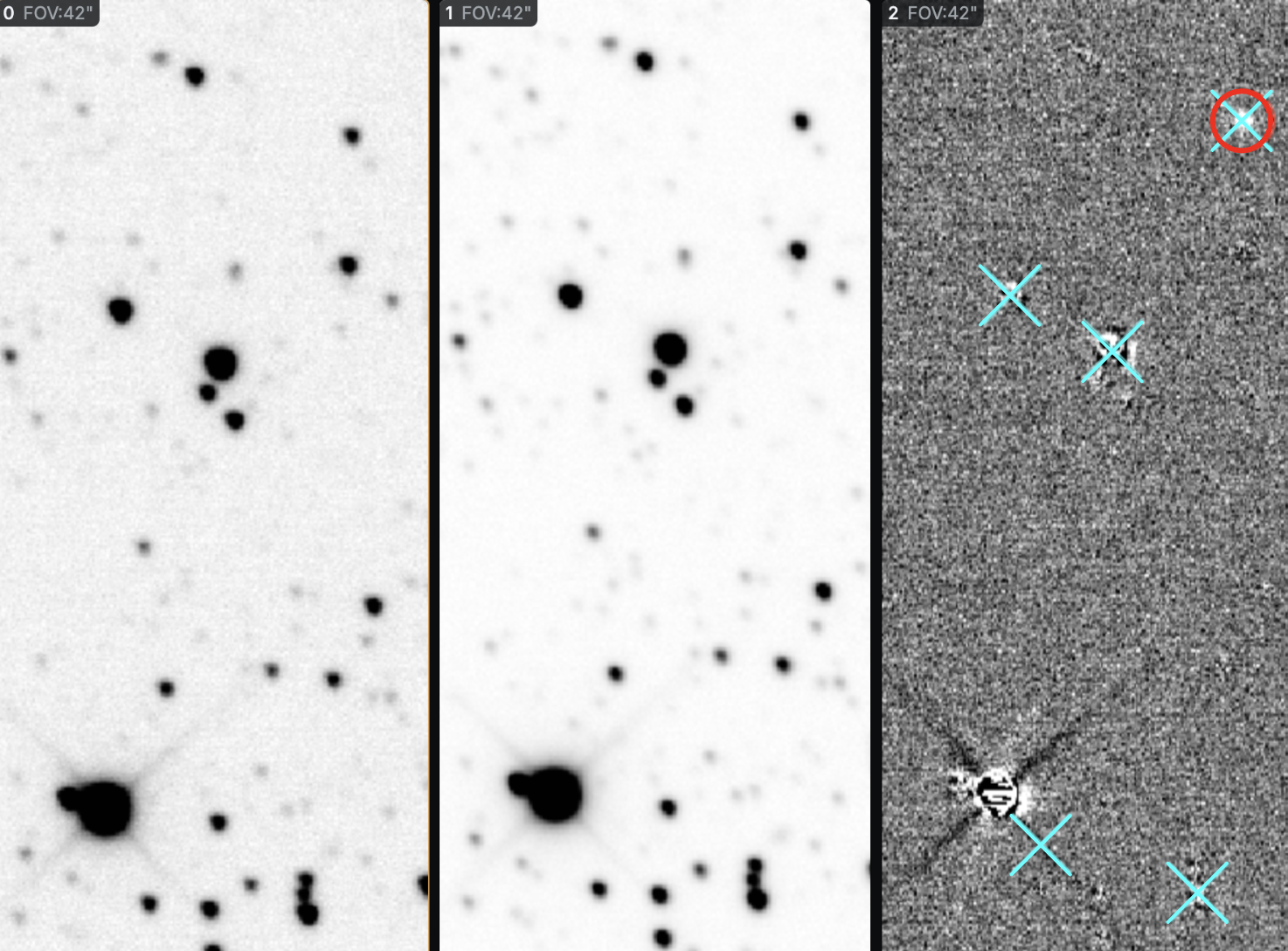}
       \caption{Science, template, and difference images of one small part of the 47 Tuc field. The cyan crosses mark \texttt{DiaSource} detections and the red circle marks a known RR Lyrae star. In these images the darker pixels correspond to higher flux and lighter pixels to lower flux. In particular, white pixels in the difference image correspond to negative flux.
      \label{fig:diaSources}}
\end{figure}

\begin{figure*}
 \centering
      \includegraphics[width=\linewidth]{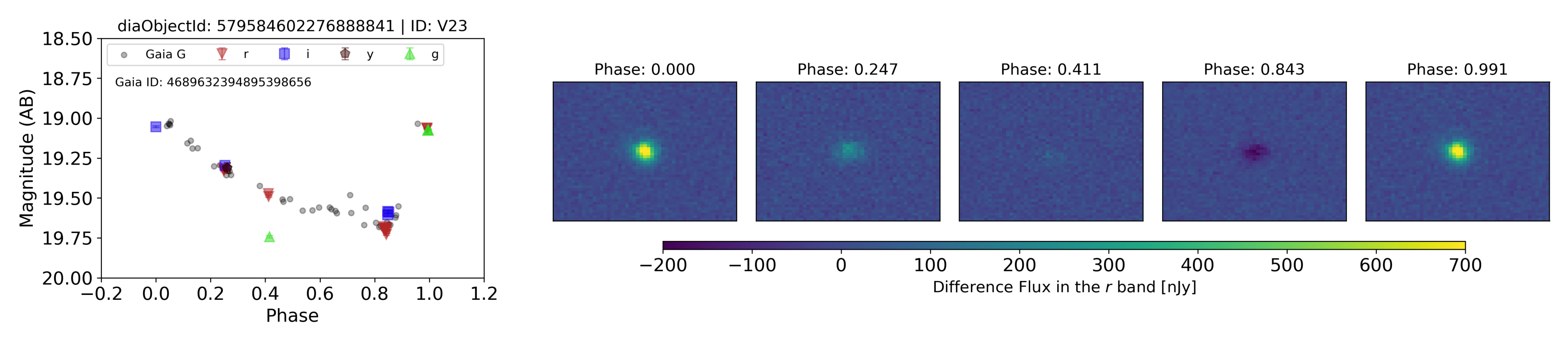}
       \caption{\textit{Left:} Phase-folded lightcurve of an SMC RR Lyrae star (OGLE SMC-RRLYR-45) recovered in our crossmatch. The \texttt{diaObjectId} and the \cite{Weldrake2004} ID are given in the title. The LSST lightcurves are based on forced PSF photometry from individual visit images (\texttt{psfFlux}). Gaia $G$-band photometry is overplotted in gray for comparison. While the LSST lightcurves are sparse given four nights of observations in the field, they still follow the Gaia lightcurve very well. \textit{Right:} Example $r$-band difference images (i.e., science image minus template) at selected phases. Each panel shows a cutout centered on the target RR Lyrae star, illustrating the variability across the pulsation cycle as recovered from difference imaging. Positive (negative) flux indicates that the star is brighter (fainter) relative to the template.
      \label{fig:V23_lc}}
\end{figure*}

In Figure~\ref{fig:diaSources}, we show the science, template, and difference images in a small field within 47 Tuc from visit \texttt{2024112400101} and detector \texttt{3}. The overplotted cyan crosses show the ``DiaSources" detected in the difference image. The source marked with the red circle is a known RR Lyrae variable star and shows a clear negative flux in the difference image. It appears some of the DiaSources may be incomplete subtractions of the brightest stars. At least one of the detections appears to be on a diffraction spike of the brightest star in the image. 

There are several flux measurements available in the DP1 datasets. The \texttt{scienceFlux} column from the \texttt{DiaSource} table is the forced PSF flux measured at the position of the DiaSource in all visit images where the DiaSource is detected. We also have forced photometry measurements at the position of the associated DiaObject on the difference image (\texttt{psfDiffFlux} column) and the visit image (\texttt{psfFlux} column) available from the \texttt{ForcedSourceOnDiaObject} table. 

In Figure~\ref{fig:V23_lc}, we plot one of the RRLyr stars (\texttt{diaObjectId} = 579584602276888841; ID: V23) recovered in our crossmatch. Because the 47 Tuc field was only visited on 4 days (Table~\ref{tab:47tuc_filters}), the lightcurve is quite sparse. However, the shape of the lightcurve looks like an RR Lyrae profile and matches the phase-wrapped lightcurve of the same source from \citet{Weldrake2004}. We plotted the forced photometry on the visit images at the DiaObject position (\texttt{psfFlux}), as there are more flux measurements in the \texttt{ForcedSourceOnDiaObject} table than in the \texttt{DiaSource} table. This is because the source is not always detected in the difference images, for example when its science flux is close to the flux in the template image, and thus does not always make it into the \texttt{DiaSource} table. The phase-folded lightcurve shows a consistent pattern with the densely-sampled one from Gaia DR3 photometry \citep{Clementini2023}.

\section{Summary}\label{sec:summary}
This study presents a detailed exploration and constructive assessment of Rubin Observatory DP1 data centered on the globular cluster 47 Tuc. The LSSTComCam imaging provided valuable early photometric measurements, while also revealing challenges from crowding, particularly the inner region of 47 Tuc and toward the SMC. Although crowding affected deblending and subsequent measurements, improvements to \texttt{scarlet} lite are anticipated to enhance processing capabilities in crowded regions before the first data release. 

We identified 3,576 high-confidence 47 Tuc member stars through a rigorous selection process combining isochrone-based filtering on the ($g - r$, $r$) color-magnitude diagram, Gaia DR3 proper-motion filtering, and multi-dimensional color-color filtering. By fitting fifth-degree polynomials to the observed stellar locus in three unique color-color diagrams, we achieved better star-galaxy separation than the provided morphology-based extendedness parameter. We also characterized the saturation limits, with caveats in the DP1 photometry in the 47 Tuc field. Further analysis of photometric scatter suggested possible contributions from differential extinction or observational systematics, indicating areas for future investigation with more data.

Further, we successfully crossmatched known variable stars within the 47 Tuc field against the DP1 data, recovering three RR Lyrae stars and two eclipsing binaries in the coadd-based object catalog, and identify 62 of the 72 known variables in the difference image-based object catalog. Despite sparse temporal sampling, forced photometry at their fixed locations in every image aligns pretty well with the denser lightcurves from the literature, underscoring Rubin's promising capabilities for variability detections even in crowded regions.  

Overall, while challenges remain, the DP1 data around 47 Tuc convincingly showcase Rubin Observatory's strong potential for detailed stellar population and variability analyses in crowded stellar fields. Continued improvements to the Rubin Science Pipelines and in-kind programs dedicated to crowded-field stellar photometry are expected to deliver even higher-quality results in future DP2 and DR1. The substantial increase in depth and cadence over the 10-yr LSST survey will enable more complete variability studies and high-fidelity characterization of stellar populations in 47 Tuc and similar dense fields.

\begin{acknowledgments}
We are grateful to the referee for their review and constructive feedback. This material is based upon work supported in part by the National Science Foundation (NSF) through Cooperative Agreement AST-1258333 and Cooperative Support Agreement AST-1202910 managed by the Association of Universities for Research in Astronomy (AURA), and the Department of Energy under Contract No. DE-AC02-76SF00515 with the SLAC National Accelerator Laboratory managed by Stanford University. Additional Rubin Observatory funding comes from private donations, grants to universities, and in-kind support from LSSTC Institutional Members. The work of Y. Choi, K. Olsen, J. Carlin, and A. Saha are supported by NSF NOIRLab, which is managed by AURA under a cooperative agreement with the U.S. National Science Foundation. 
\end{acknowledgments}

\facilities{Rubin:Simonyi (LSSTComCam), USDAC, USDF}

\software{\texttt{LSST Science Pipelines} \citep{PSTN-019}, \texttt{Qserv} \citep{10.1145/2063348.2063364, DMTN-243}, \texttt{Rubin Data Butler}\citep{2022SPIE12189E..11J}, \texttt{Astropy} \citep{Astropy2013, Astropy2018, Astropy2022}, \texttt{Matplotlib} \citep{Hunter2007}, \texttt{NumPy} \citep{vanderwalt2011, harris2020}, \texttt{scikit-learn} \citep{Pedregosa2011}, \texttt{SciPy} \citep{Scipy2020}.
          }

\end{document}

%% file: authors.tex
\author[0000-0003-1680-1884]{Yumi Choi}
\affiliation{NSF National Optical-Infrared Astronomy Research Laboratory, 950 North Cherry Avenue, Tucson, AZ 85719, USA}

\author[0000-0002-7134-8296]{Knut A. G. Olsen}
\affiliation{NSF National Optical-Infrared Astronomy Research Laboratory, 950 North Cherry Avenue, Tucson, AZ 85719, USA}

\author[0000-0002-3936-9628]{Jeffrey L. Carlin}
\affiliation{NSF NOIRLab/NSF–DOE Vera C. Rubin Observatory HQ, 950 N. Cherry Ave., Tucson, AZ 85719, USA}

\author[0000-0001-5538-0395]{Yuankun (David) Wang}
\affiliation{DIRAC Institute, Department of Astronomy, University of Washington, 3910 15th Avenue NE, Seattle, WA 98195, USA}

\author[0000-0003-0093-4279]{Fred Moolekamp}
\affiliation{soZen Inc., 105 Clearview Dr, Penfield, NY 14526}

\author[0000-0002-6839-4881]{Abhijit Saha}
\affiliation{NSF National Optical-Infrared Astronomy Research Laboratory, 950 North Cherry Avenue, Tucson, AZ 85719, USA}

\author{Ian Sullivan}
\affiliation{DiRAC Institute, Department of Astronomy, University of Washington, Seattle, WA 98195, USA; Vera C. Rubin Observatory, Chile}

\author[0000-0002-0558-0521]{Colin T. Slater}
\affiliation{Department of Astronomy, University of Washington, Box 351580, Seattle, WA 98195-1580, USA}

\author[0000-0001-7211-5729]{Douglas L. Tucker}
\affiliation{Fermi National Accelerator Laboratory, P. O. Box 500, Batavia, IL 60510, USA}

\author[0009-0008-8623-871X]{Christina L. Adair}
\affiliation{SLAC National Accelerator Laboratory, Stanford University, 2575 Sand Hill Road, Menlo Park, CA 94025, USA}

\author[0000-0001-6957-1627]{Peter S. Ferguson}
\affiliation{DiRAC Institute, Department of Astronomy, University of Washington, 3910 15th Ave NE, Seattle, WA, 98195, USA}

\author[0000-0002-5261-5803]{Yijung Kang}
\affiliation{Kavli Institute for Particle Astrophysics and Cosmology, SLAC National Accelerator Laboratory, Stanford University, 2575 Sand Hill Road, Menlo Park, CA 94025, USA}

\author[0000-0002-5855-401X]{Karla Peña Ramírez}
\affiliation{NSF NOIRLab/Vera C. Rubin Observatory, Casilla 603, La Serena, Chile}

\author[0000-0003-2935-7196]{Markus Rabus}
\affiliation{Departamento de Matem{\'a}tica y F{\'i}sica Aplicadas, Facultad de Ingenier{\'i}a, Universidad Cat{\'o}lica de la Sant{\'i}sima Concepci{\'o}n, Alonso de Rivera 2850, Concepci{\'o}n, Chile }